\documentclass{article}
\usepackage{graphicx}
\usepackage{amsmath}

\renewcommand{\theequation}{\arabic{section}.\arabic{equation}}
\begin{document}

\title{Flavour mixing, gauge invariance and wave-function renormalisation }
\author{\textsc{D. Espriu\thanks{%
espriu@ecm.ub.es}, J. Manzano\thanks{%
manzano@ecm.ub.es}, P. Talavera\thanks{%
pere@ecm.ub.es}} \\
{Departament d'Estructura i Constituents de la Mat\`eria}\\
and \\
{CER for Astrophysics, Particle Physics and Cosmology} \\
{Universitat de Barcelona, Diagonal 647, Barcelona E-08028 Spain}\\
}
\date{}
\maketitle

\begin{abstract}
We clarify some aspects of the LSZ formalism and wave function
renormalisation for unstable particles in the presence of electroweak
interactions when mixing and $CP$ violation are considered. We also analyse
the renormalisation of the CKM mixing matrix which is closely related to
wave function renormalisation. We critically review earlier attempts to
define a set of ``on-shell'' wave function renormalisation constants. With
the aid of an extensive use of the Nielsen identities complemented by
explicit calculations we corroborate that the counter term for the CKM
mixing matrix must be explicitly gauge independent and demonstrate that the
commonly used prescription for the wave function renormalisation constants
leads to gauge parameter dependent amplitudes, even if the CKM counter term
is gauge invariant as required. We show that a proper LSZ-compliant
prescription leads to gauge independent amplitudes. The resulting wave
function renormalisation constants necessarily possess absorptive parts, but
we verify that they comply with the expected requirements concerning $CP$
and $CPT$. The results obtained using this prescription are different (even
at the level of the modulus squared of the amplitude) from the ones
neglecting the absorptive parts in the case of top decay. The difference is
numerically relevant.
\end{abstract}


\vfill
\vbox{
UB-ECM-PF 02/06\null\par
April 2002}

\section{Introduction}

\renewcommand{\theequation}{\arabic{section}.\arabic{equation}} %
\setcounter{equation}{0}

One of the pressing open problems in particle physics is to understand the
origin of $CP$ violation phase and family mixing. In the minimal Standard
Model (SM) the information about these quantities is encoded in the
Cabibbo-Kobayashi-Maskawa (CKM) mixing matrix. In this work we shall denote
this matrix by $K_{ij}$

As it is well known, some of the entries of this matrix are remarkably well
measured, while others (such as the $K_{tb}$, $K_{ts}$ and $K_{td}$
elements) are poorly known and the only real experimental constraint come
from the unitarity requirements. A lot of effort in the last decade has been
invested in this particular problem and this dedication will continue in the
foreseeable future aiming to a precision in the charged current sector
comparable to the one already reached in the neutral sector. As a guidance,
let us mention that the expected accuracy in $\sin 2\beta $ after LHCb is
expected to be beyond the 1\% level, and a comparable accuracy is expected
by that time from the ongoing generation of experiments (BaBar, Belle) \cite
{Amato:1998xt}.

In the neutral sector it is totally mandatory to include electroweak
radiative corrections to bring theory and experiment into agreement. Tree
level results are incompatible with experiment by many standard deviations
\cite{Pich:1997ga}. Obviously we are not there yet in the charged current
sector, but in a few years electroweak radiative corrections will be
required in the studies analysing the ``unitarity'' of the CKM matrix%
\footnote{%
The CKM matrix is certainly unitary, but the physical observables that at
tree level coincide with these matrix elements certainly do not necessarily
fulfil a unitarity constraint once quantum corrections are switched on.}.

These corrections are of several types. With an on-shell scheme in mind, we
need counter terms for the electric charge, Weinberg angle and wave-function
renormalisation (wfr.) for the $W$ gauge boson. We shall also require wfr.
for the external fermions and counter terms for the entries of the CKM
matrix. The latter are in fact related in a way that will be described below
\cite{Balzereit:1999id}. Finally one needs to compute the 1PI vertex parts
of the different processes one is interested in.

In the on-shell scheme, all counter terms can be expressed as combinations
of self-energies \cite{Hollik}. These are standard and well known at
one-loop in perturbation theory; in some cases, at least for the leading
pieces, up to two-loop in the SM. However, a long standing controversy
exists in the literature concerning what is the appropriate way to define
both an external wfr. and CKM counter terms. The issue becomes involved
because we are dealing with particles which are unstable (and therefore the
self-energies develop branch cuts; even gauge dependent ones in the SM) and
because of mixing.

Several proposals have been put forward in the literature to define
appropriate counter terms both for the external legs and for the CKM matrix
elements. The original prescription for wfr. diagonalizing the on-shell
propagator was introduced in \cite{Aoki}. In \cite{DennerSack} the wfr.
``satisfying'' the conditions of \cite{Aoki} were derived. However since
\cite{DennerSack} does not take care about the branch cuts present in the
self-energies those results must be considered only consistent up to
absorptive terms. Later it was realized \cite{Denner} that the on-shell
conditions defined in \cite{Aoki} where inconsistent and in fact impossible
to satisfy for a minimal set of renormalisation constants\footnote{%
By minimal set we mean a set where the wfr. of $\bar{\Psi}_{0}=\bar{\Psi}%
\bar{Z}^{\frac{1}{2}}$ and $\Psi _{0}=Z^{\frac{1}{2}}\Psi $ are related by $%
\bar{Z}^{\frac{1}{2}}=\gamma ^{0}Z^{\frac{1}{2}\dagger }\gamma ^{0}.$} due
to the imaginary branch cuts present in the self-energies. The author of
\cite{Denner} circumvented this problem by introducing a prescription that
\textit{de facto} eliminates such branch cuts, but at the price of not
diagonalizing the propagators in flavour space.

Ward identities based on the SU(2)$_{L}$ gauge symmetry relate wfr. and
counter terms for the CKM matrix elements \cite{Balzereit:1999id}. In \cite
{Grassi} it was seen that if the prescription of \cite{DennerSack} was used
in the counter terms for the CKM matrix elements, the results were in
violation of gauge invariance. As we have just mentioned, the results in
\cite{DennerSack} do not deal properly with the absorptive terms appearing
in the self-energies; which in addition happen to be gauge dependent. In
spite of the problems with the prescription for the wfr. given in \cite
{DennerSack}, the conclusions reached in \cite{Grassi} are correct: a
necessary condition for gauge invariance of the physical amplitudes is that
counter terms for the CKM matrix elements $K_{ij}$ are by themselves gauge
independent. This condition is fulfilled by the CKM counter term proposed in
\cite{Grassi} as it is in minimal subtraction \cite{Balzereit:1999id}, \cite
{Diener:2001qt}.

Other proposals to handle CKM renormalisation exist in the literature \cite
{Diener:2001qt}, \cite{Barroso} and \cite{Yamada}. In all these works either
the external wfr. proposed originally in \cite{DennerSack} or \cite{Denner}
are used, or the issue is sidestepped altogether. In either case the
absorptive part of the self-energies (and even the absorptive part of the
1PI vertex part in one particular instance \cite{Barroso}) are not taken
into account. As we shall see doing so leads to physical amplitudes --- $S$%
-matrix elements--- which are gauge dependent, and this irrespective of the
method one uses to renormalise $K_{ij}$ provided the redefinition of $K_{ij}$
is gauge independent and preserves unitarity.

Due to the structure of the imaginary branch cuts it turns out however, that
the gauge dependence present in the amplitude using the prescription of \cite
{Denner} cancels in the modulus squared of the physical $S$-matrix element
in the SM. This cancellation has been checked numerically by the authors in
\cite{Kniehl}. In this work we shall provide analytical results showing that
this cancellation is exact. However the gauge dependence remains at the
level of the amplitude.

Is this acceptable? We do not think so. Diagrams contributing to the same
physical process outside the SM electroweak sector may interfere with the SM
amplitude and reveal the unwanted gauge dependence. Furthermore, gauge
independent absorptive parts are also discarded by the prescription in \cite
{Denner}. These parts, contrary to the gauge dependent ones, do not drop in
the squared amplitude as we shall show. In addition, one should not forget
that the scheme in \cite{Denner} does not deliver on-shell renormalised
propagators that are diagonal in flavour space.

This work is dedicated to substantiate the above claims. We shall compute
the gauge dependence of the absorptive parts in the self-energies and the
vertex functions. We shall see how the requirements of gauge invariance and
proper on-shell conditions (including exact diagonalisation in flavour
space) single out a unique prescription for the wfr. The problem is
presented in detail in the next section. The explicit expressions for the
renormalisation constants are given in sections \ref{offdiagonal} and \ref
{diagonal}. Implementation for $W$ and top decay are shown in section \ref
{wandtdecay}. A technical discussion where extended use of the Nielsen
identities has been done to extract the gauge dependence of all absorptive
terms is presented in section \ref{nielsen} and it can be omitted by readers
not interested in these details. In section \ref{absorptive} and \ref{cp} we
return to $W$ and top decay to implement the previous results and finally we
conclude in section \ref{conc}.

\section{Statement of the problem and its solution}

\renewcommand{\theequation}{\arabic{section}.\arabic{equation}} %
\setcounter{equation}{0}

We want to define an on-shell renormalisation scheme that guarantees the
correct properties of the fermionic propagator in the $p^{2}\rightarrow
m_{i}^{2}$ limit and at the same time renders the observable quantities
calculated in such a scheme gauge parameter independent. In the first place
up and down-type propagators have to be family diagonal on-shell. The
conditions necessary for that purpose were first given by Aoki et. al. in
\cite{Aoki}. Let us introduce some notation in order to write them down. We
renormalise the bare fermion fields $\Psi _{0}$ and $\bar{\Psi}_{0}$ as
\begin{equation}
\Psi _{0}=Z^{\frac{1}{2}}\Psi \,,\qquad \bar{\Psi}_{0}=\bar{\Psi}\bar{Z}^{%
\frac{1}{2}}\,.  \label{fundamental}
\end{equation}
For reasons that will become clear along the discussion, we shall allow $Z$
and $\bar{Z}$ to be independent renormalisation constants\footnote{%
This immediately raises some issues about hermiticity which we shall deal
with below.}. These renormalisation constants contain flavour, family and
Dirac indices. We can decompose them into
\begin{equation}
Z^{\frac{1}{2}}=Z^{u\frac{1}{2}}\tau ^{u}+Z^{d\frac{1}{2}}\tau ^{d}\,,\qquad
\bar{Z}^{\frac{1}{2}}=\bar{Z}^{u\frac{1}{2}}\tau ^{u}+\bar{Z}^{d\frac{1}{2}%
}\tau ^{d}\,,  \label{ud}
\end{equation}
with $\tau ^{u}$ and $\tau ^{d}$ the up and down flavour projectors and
furthermore each piece in left and right chiral projectors, $L$ and $R$
respectively,
\begin{equation}
Z^{u\frac{1}{2}}=Z^{uL\frac{1}{2}}L+Z^{uR\frac{1}{2}}R\,,\qquad \bar{Z}^{u%
\frac{1}{2}}=\bar{Z}^{uL\frac{1}{2}}R+\bar{Z}^{uR\frac{1}{2}}L\,.  \label{LR}
\end{equation}
Analogous decompositions hold for $Z^{d\frac{1}{2}}$ and $\bar{Z}^{d\frac{1}{%
2}}$. Due to radiative corrections the propagator mixes fermion of different
family indices. Namely
\begin{equation*}
iS^{-1}\left( p\right) =\bar{Z}^{\frac{1}{2}}\left( \frac{{}}{{}}\not{p}%
-m-\delta m-\Sigma \left( p\right) \right) Z^{\frac{1}{2}}\,,
\end{equation*}
where the bare self-energy $\Sigma $ is non-diagonal and is given by $%
-i\Sigma =\sum $1PI. Within one-loop accuracy we can write $Z^{\frac{1}{2}%
}=1+\frac{1}{2}\delta Z$ etc. Introducing the family indices explicitly we
have
\begin{equation*}
iS_{ij}^{-1}\left( p\right) =\left( \not{p}-m_{i}\right) \delta _{ij}-\hat{%
\Sigma}_{ij}\left( p\right) \,,
\end{equation*}
where the one-loop renormalised self-energy is given by
\begin{equation}
\hat{\Sigma}_{ij}\left( p\right) =\Sigma _{ij}\left( p\right) -\frac{1}{2}%
\delta \bar{Z}_{ij}\left( \not{p}-m_{j}\right) -\frac{1}{2}\left( \not{p}%
-m_{i}\right) \delta Z_{ij}+\delta m_{i}\delta _{ij}\,.  \label{renself}
\end{equation}
Since we can project the above definition for up and down type-quarks,
flavour indices will be dropped in the sequel and only will be restored when
necessary. Recalling the following on-shell relations for Dirac spinors ($%
p^{2}\rightarrow m_{i}^{2}$)
\begin{eqnarray}
\left( \not{p}-m_{i}\right) u_{i}^{\left( s\right) }\left( p\right) &=&0\,,
\notag \\
\bar{u}_{i}^{\left( s\right) }\left( p\right) \left( \not{p}-m_{i}\right)
&=&0\,,  \notag \\
\left( \not{p}-m_{i}\right) v_{i}^{\left( s\right) }\left( -p\right) &=&0\,,
\notag \\
\bar{v}_{i}^{\left( s\right) }\left( -p\right) \left( \not{p}-m_{i}\right)
&=&0\,,  \label{onshellspinors}
\end{eqnarray}
the conditions \cite{Aoki} necessary to avoid mixing will be\footnote{%
Notice that, as a matter of fact, in \cite{Aoki} the conditions over
anti-fermions are not stated.}

\begin{eqnarray}
\hat{\Sigma}_{ij}\left( p\right) u_{j}^{\left( s\right) }\left( p\right)
&=&0\,,\qquad (p^{2}\rightarrow m_{j}^{2})\,,\quad \mathrm{(incoming}\text{ }%
\mathrm{particle)}  \label{inparticle} \\
\bar{v}_{i}^{\left( s\right) }\left( -p\right) \hat{\Sigma}_{ij}\left(
p\right) &=&0\,,\qquad (p^{2}\rightarrow m_{i}^{2})\,,\quad \mathrm{(incoming%
}\text{ }\mathrm{anti}\mathrm{-}\mathrm{particle)}  \label{inantiparticle} \\
\bar{u}_{i}^{\left( s\right) }\left( p\right) \hat{\Sigma}_{ij}\left(
p\right) &=&0\,,\qquad (p^{2}\rightarrow m_{i}^{2})\,,\quad \mathrm{(outgoing%
}\text{ }\mathrm{particle)}  \label{outparticle} \\
\hat{\Sigma}_{ij}\left( p\right) v_{j}^{\left( s\right) }\left( -p\right)
&=&0\,,\qquad (p^{2}\rightarrow m_{j}^{2})\,,\quad \mathrm{(outgoing}\text{ }%
\mathrm{anti}\mathrm{-}\mathrm{particle)}  \label{outantiparticle}
\end{eqnarray}
where no summation over repeated indices is assumed and $i\neq j.$ These
relations determine the non-diagonal parts of $Z$ and $\bar{Z}$ as will be
proven in the next section. Here, as a side remark, let us point out that
the need of different ``incoming'' and ``outgoing'' wfr. constants was already
recognised in \cite{espriumanz}. Nevertheless, that paper was unsuccessful
in reconciling the on-shell prescription with the presence of absorptive
terms in the self-energies. However, since its results are concerned with
the leading contribution of an effective Lagrangian, no absorptive terms are
present and therefore conclusions still hold.

To obtain the diagonal parts $Z_{ii},$ $\bar{Z}_{ii}$, and $\delta m_{i}$
one imposes mass pole and unit residue conditions (to be discussed below).
Here it is worth to make one important comment regarding the above
conditions. First of all we note that in the literature the relation
\begin{equation}
\bar{Z}^{\frac{1}{2}}=\gamma ^{0}Z^{\frac{1}{2}\dagger }\gamma ^{0}\,,
\label{hermiticity}
\end{equation}
is taken for granted. This relation is tacitly assumed in \cite{Aoki} and
explicitly required in \cite{Denner}. Imposing Eq. (\ref{hermiticity}) would
guarantee hermiticity of the Lagrangian written in terms of the renormalised
physical fields. However, we are at this point concerned with external leg
renormalisation, for which it is perfectly possible to use a different set
of renormalisation constants (even ones that do not respect the requirement (%
\ref{hermiticity})), while keeping the Lagrangian hermitian. In fact, using
two sets of renormalisation constants is a standard practice in the on-shell
scheme \cite{Hollik}, so one should not be concerned by this fact \textit{%
per se}. In case one is worried about the consistency of using a set of wfr.
constants not satisfying (\ref{hermiticity}) for the external legs while
keeping a hermitian Lagrangian, it should be pointed out that there is a
complete equivalence between the set of renormalisation constants we shall
find out below and a treatment of the external legs where diagrams with
self-energies (including mass counter terms) are inserted instead of the
wfr. constants; provided, of course, that the mass counter term satisfy the
on-shell condition. Proceeding in this way gives results identical to ours
and different from those obtained using the wfr. proposed in \cite{Denner},
which do fulfil (\ref{hermiticity}). Further consistency checks are
presented in the following sections.

In any case, self-energies develop absorptive terms and this makes Eq. (\ref
{hermiticity}) incompatible with the diagonalizing conditions (\ref
{inparticle})-(\ref{outantiparticle}). Therefore in order to circumvent this
problem one can give up diagonalisation conditions (\ref{inparticle})-(\ref
{outantiparticle}) or alternatively the hermiticity condition (\ref
{hermiticity}). The approach taken originally in \cite{Denner} and works
thereafter was the former alternative, while in this work we shall advocate
the second one. The approach of \cite{Denner} consists in dropping out
absorptive terms from conditions (\ref{inparticle})-(\ref{outantiparticle}).
That is for $i\neq j$
\begin{eqnarray}
\widetilde{Re}\left( \hat{\Sigma}_{ij}\left( p\right) \right) u_{j}^{\left(
s\right) }\left( p\right) &=&0\,,\qquad (p^{2}\rightarrow m_{j}^{2})\,,\quad
\mathrm{(incoming}\text{ }\mathrm{particle)}  \notag \\
\bar{v}_{i}^{\left( s\right) }\left( -p\right) \widetilde{Re}\left( \hat{%
\Sigma}_{ij}\left( p\right) \right) &=&0\,,\qquad (p^{2}\rightarrow
m_{i}^{2})\,,\quad \mathrm{(incoming}\text{ }\mathrm{anti}\mathrm{-}\mathrm{%
particle)}  \notag \\
\bar{u}_{i}^{\left( s\right) }\left( p\right) \widetilde{Re}\left( \hat{%
\Sigma}_{ij}\left( p\right) \right) &=&0\,,\qquad (p^{2}\rightarrow
m_{i}^{2})\,,\quad \mathrm{(outgoing}\text{ }\mathrm{particle)}  \notag \\
\widetilde{Re}\left( \hat{\Sigma}_{ij}\left( p\right) \right) v_{j}^{\left(
s\right) }\left( -p\right) &=&0\,,\qquad (p^{2}\rightarrow
m_{j}^{2})\,,\quad \mathrm{(outgoing}\text{ }\mathrm{anti}\mathrm{-}\mathrm{%
particle)}  \label{Denner}
\end{eqnarray}
where $\widetilde{Re}$ includes the real part of the logarithms arising in
loop integrals appearing in the self-energies but not of the rest of
coupling factors of the Feynmann diagram. This approach is compatible with
the hermiticity condition (\ref{hermiticity}) but on the other hand have
several drawbacks. These drawbacks include

\begin{enumerate}
\item  Since only the $\widetilde{Re}$ part of the self-energies enters into
the diagonalizing conditions the on-shell propagator remains non-diagonal.

\item  The very definition of $\widetilde{Re}$ relies heavily on the
one-loop perturbative calculation where it is applied upon. In other words $%
\widetilde{Re}$ is not a proper function of its argument (in contrast to $Re$%
) and it is presumably cumbersome to implement in multi-loop calculations.

\item  As it will become clear in next sections, the on-shell scheme based
in the $\widetilde{Re}$ prescription leads to gauge parameter dependent
physical amplitudes. The reason for this unwanted dependence is the dropping
of absorptive gauge parameter dependent terms in the self-energies that are
necessary to cancel absorptive terms appearing in the vertices. As mentioned
in the introduction, in the SM, the gauge dependence drops in the modulus
squared of the amplitude, but not in the amplitude itself and it could be
eventually observable.
\end{enumerate}

Once stated the unwanted features of the $\widetilde{Re}$ approach let us
briefly state the consequences of dropping condition (\ref{hermiticity})

\begin{enumerate}
\item  Conditions (\ref{inparticle})-(\ref{outantiparticle}) readily
determine the off-diagonal $Z$ and $\bar{Z}$ wfr. which coincide with the
ones obtained using the $\widetilde{Re}$ prescription up to finite
absorptive gauge parameter dependent terms.

\item  The renormalised fermion propagator becomes exactly diagonal
on-shell, unlike in the $\widetilde{Re}$ scheme.

\item  Incoming and outgoing particles and anti-particles require different
renormalisation constants when computing a physical amplitude. Annihilation
of particles and creation of anti-particles are accompanied by the
renormalisation constant $Z$, while creation of particles and annihilation
of anti-particles are accompanied by the renormalisation constant $\bar{Z}$.

\item  These constants $Z$ and $\bar{Z}$ are in what respects to their
dispersive parts identical to the ones in \cite{Denner}. They differ in
their absorptive parts. This might suggest to the alert reader there could
be problems with fundamental symmetries such as $CP$ or $CPT$. We shall
discuss this issue at the end of the paper. Our conclusion is that
everything works out consistently in this respect.
\end{enumerate}

For explicit expressions for $Z$ and $\bar{Z}$ the reader should consult
formulae (\ref{zin}), (\ref{zout}) and (\ref{zdiag}) in the next two
sections. As an example how to implement them see section \ref{wandtdecay}.
The explicit dependence on the gauge parameter (for simplicity only the $W$
gauge parameter is considered) of the absorptive parts is given in section
\ref{absorptive}.

\section{Off-diagonal wave-function renormalisation constants}

\renewcommand{\theequation}{\arabic{section}.\arabic{equation}} %
\setcounter{equation}{0}

\label{offdiagonal}This section is devoted to a detailed derivation of the
off-diagonal renormalisation constants deriving entirely from the on-shell
conditions (\ref{inparticle})-(\ref{outantiparticle}) and allowing for $\bar{%
Z}^{\frac{1}{2}}\neq \gamma ^{0}Z^{\frac{1}{2}\dagger }\gamma ^{0}$. First
of all we decompose the renormalised self-energy into all possible Dirac
structures

\begin{equation}
\hat{\Sigma}_{ij}\left( p\right) =\not{p}\left( \hat{\Sigma}_{ij}^{\gamma
R}\left( p^{2}\right) R+\hat{\Sigma}_{ij}^{\gamma L}\left( p^{2}\right)
L\right) +\hat{\Sigma}_{ij}^{R}\left( p^{2}\right) R+\hat{\Sigma}%
_{ij}^{L}\left( p^{2}\right) L\,,  \label{decomp}
\end{equation}
and use Eqs. (\ref{LR}), (\ref{renself}) and (\ref{decomp}) to obtain
\begin{eqnarray}
\hat{\Sigma}_{ij}\left( p\right) &=&\not{p}R\left( \Sigma _{ij}^{\gamma
R}\left( p^{2}\right) -\frac{1}{2}\delta \bar{Z}_{ij}^{R}-\frac{1}{2}\delta
Z_{ij}^{R}\right) +\not{p}L\left( \Sigma _{ij}^{\gamma L}\left( p^{2}\right)
-\frac{1}{2}\delta \bar{Z}_{ij}^{L}-\frac{1}{2}\delta Z_{ij}^{L}\right)
\notag  \label{renormself} \\
&&\hspace{-1.3cm}+R\left( \Sigma _{ij}^{R}\left( p^{2}\right) +\frac{1}{2}%
\left( \delta \bar{Z}_{ij}^{L}m_{j}+m_{i}\delta Z_{ij}^{R}\right) +\delta
_{ij}\delta m_{i}\right) +L\left( \Sigma _{ij}^{L}\left( p^{2}\right) +\frac{%
1}{2}\left( \delta \bar{Z}_{ij}^{R}m_{j}+m_{i}\delta Z_{ij}^{L}\right)
+\delta _{ij}\delta m_{i}\right) \,.  \notag \\
&&
\end{eqnarray}
Repeated indices are not summed over. Hence from Eqs. (\ref{renormself}), (%
\ref{onshellspinors}) and (\ref{inparticle}) we obtain
\begin{eqnarray*}
\Sigma _{ij}^{\gamma R}\left( m_{j}^{2}\right) m_{j}-\frac{1}{2}\delta
Z_{ij}^{R}m_{j}+\Sigma _{ij}^{L}\left( m_{j}^{2}\right) +\frac{1}{2}%
m_{i}\delta Z_{ij}^{L} &=&0\,, \\
\Sigma _{ij}^{\gamma L}\left( m_{j}^{2}\right) m_{j}-\frac{1}{2}\delta
Z_{ij}^{L}m_{j}+\Sigma _{ij}^{R}\left( m_{j}^{2}\right) +\frac{1}{2}%
m_{i}\delta Z_{ij}^{R} &=&0\,.
\end{eqnarray*}
Exactly the same relations are obtained from Eqs. (\ref{renormself}), (\ref
{onshellspinors}) and Eq. (\ref{outantiparticle}). Analogously, Eqs. (\ref
{renormself}), (\ref{onshellspinors}) and Eq. (\ref{inantiparticle}) (or Eq.
(\ref{outparticle})) lead to
\begin{eqnarray*}
m_{i}\Sigma _{ij}^{\gamma R}\left( m_{i}^{2}\right) -\frac{1}{2}m_{i}\delta
\bar{Z}_{ij}^{R}+\Sigma _{ij}^{R}\left( m_{i}^{2}\right) +\frac{1}{2}\delta
\bar{Z}_{ij}^{L}m_{j} &=&0\,, \\
m_{i}\Sigma _{ij}^{\gamma L}\left( m_{i}^{2}\right) -\frac{1}{2}m_{i}\delta
\bar{Z}_{ij}^{L}+\Sigma _{ij}^{L}\left( m_{i}^{2}\right) +\frac{1}{2}\delta
\bar{Z}_{ij}^{R}m_{j} &=&0\,.
\end{eqnarray*}
Using the above expressions we immediately obtain
\begin{eqnarray}
\delta Z_{ij}^{L} &=&\frac{2}{m_{j}^{2}-m_{i}^{2}}\left[ \Sigma
_{ij}^{\gamma R}\left( m_{j}^{2}\right) m_{i}m_{j}+\Sigma _{ij}^{\gamma
L}\left( m_{j}^{2}\right) m_{j}^{2}+m_{i}\Sigma _{ij}^{L}\left(
m_{j}^{2}\right) +\Sigma _{ij}^{R}\left( m_{j}^{2}\right) m_{j}\right] \,,
\notag \\
\delta Z_{ij}^{R} &=&\frac{2}{m_{j}^{2}-m_{i}^{2}}\left[ \Sigma
_{ij}^{\gamma L}\left( m_{j}^{2}\right) m_{i}m_{j}+\Sigma _{ij}^{\gamma
R}\left( m_{j}^{2}\right) m_{j}^{2}+m_{i}\Sigma _{ij}^{R}\left(
m_{j}^{2}\right) +\Sigma _{ij}^{L}\left( m_{j}^{2}\right) m_{j}\right] \,,
\label{zin}
\end{eqnarray}
and
\begin{eqnarray}
\delta \bar{Z}_{ij}^{L} &=&\frac{2}{m_{i}^{2}-m_{j}^{2}}\left[ \Sigma
_{ij}^{\gamma R}\left( m_{i}^{2}\right) m_{i}m_{j}+\Sigma _{ij}^{\gamma
L}\left( m_{i}^{2}\right) m_{i}^{2}+m_{i}\Sigma _{ij}^{L}\left(
m_{i}^{2}\right) +\Sigma _{ij}^{R}\left( m_{i}^{2}\right) m_{j}\right] \,,
\notag \\
\delta \bar{Z}_{ij}^{R} &=&\frac{2}{m_{i}^{2}-m_{j}^{2}}\left[ \Sigma
_{ij}^{\gamma L}\left( m_{i}^{2}\right) m_{i}m_{j}+\Sigma _{ij}^{\gamma
R}\left( m_{i}^{2}\right) m_{i}^{2}+m_{i}\Sigma _{ij}^{R}\left(
m_{i}^{2}\right) +\Sigma _{ij}^{L}\left( m_{i}^{2}\right) m_{j}\right] \,.
\label{zout}
\end{eqnarray}
At the one-loop level in the SM we can define
\begin{equation*}
\Sigma _{ij}^{R}\left( p^{2}\right) \equiv \Sigma _{ij}^{S}\left(
p^{2}\right) m_{j}\,,\qquad \Sigma _{ij}^{L}\left( p^{2}\right) \equiv
m_{i}\Sigma _{ij}^{S}\left( p^{2}\right) \,,
\end{equation*}
and therefore
\begin{eqnarray*}
\delta \bar{Z}_{ij}^{L}-\delta Z_{ij}^{L\dagger } &=&\frac{2}{%
m_{i}^{2}-m_{j}^{2}}\left\{ \frac{{}}{{}}\left( \Sigma _{ij}^{\gamma
R}\left( m_{i}^{2}\right) -\Sigma _{ji}^{\gamma R\ast }\left(
m_{i}^{2}\right) \right) m_{i}m_{j}\right. +\left( \Sigma _{ij}^{\gamma
L}\left( m_{i}^{2}\right) -\Sigma _{ji}^{\gamma L\ast }\left(
m_{i}^{2}\right) \right) m_{i}^{2} \\
&&\left. +\left( m_{i}^{2}+m_{j}^{2}\right) \left( \frac{{}}{{}}\Sigma
_{ij}^{S}\left( m_{i}^{2}\right) -\Sigma _{ji}^{S\ast }\left(
m_{i}^{2}\right) \frac{{}}{{}}\right) \frac{{}}{{}}\right\} \neq 0\,,
\end{eqnarray*}
and a similar relation holds for $\delta \bar{Z}_{ij}^{R}-\delta
Z_{ij}^{R\dagger }.$ The above non-vanishing difference is due to the
presence of branch cuts in the self-energies that invalidate the
pseudo-hermiticity relation
\begin{equation}
\Sigma _{ij}\left( p\right) \neq \gamma ^{0}\Sigma _{ij}^{\dagger }\left(
p\right) \gamma ^{0}\,.  \label{pseudo}
\end{equation}
Eq. (\ref{pseudo}) is assumed in \cite{Aoki} and if we, temporally, ignore
those branch cut contributions our results reduces to the ones depicted in
\cite{DennerSack} or \cite{Denner}. In the SM these branch cuts are
generically gauge dependent as a cursory look to the appropriate integrals
shows at once.

\section{Diagonal wave-function renormalisation constants}

\renewcommand{\theequation}{\arabic{section}.\arabic{equation}} %
\setcounter{equation}{0}

\label{diagonal} Once the off-diagonal wfr. are obtained we focus our
attention in the diagonal sector. Near the on-shell limit we can neglect the
off-diagonal parts of the inverse propagator and write
\begin{equation}
iS_{ij}^{-1}\left( p\right) =\left( \not{p}-m_{i}-\hat{\Sigma}_{ii}\left(
p\right) \right) \delta _{ij}=\left( \frac{{}}{{}}\not{p}\left( aL+bR\right)
+cL+dR\frac{{}}{{}}\right) \delta _{ij}\,,
\end{equation}
and therefore after some algebra
\begin{equation*}
-iS_{ij}\left( p\right) =\frac{\not{p}\left( aL+bR\right) -dL-cR}{p^{2}ab-cd}%
\delta _{ij}\,,
\end{equation*}
in our case we have
\begin{eqnarray}
a &=&1-\Sigma _{ii}^{\gamma L}\left( p^{2}\right) +\frac{1}{2}\delta \bar{Z}%
_{ii}^{L}+\frac{1}{2}\delta Z_{ii}^{L}\,,  \notag \\
b &=&1-\Sigma _{ii}^{\gamma R}\left( p^{2}\right) +\frac{1}{2}\delta \bar{Z}%
_{ii}^{R}+\frac{1}{2}\delta Z_{ii}^{R}\,,  \notag \\
c &=&-\Sigma _{ii}^{L}\left( p^{2}\right) -\left( 1+\frac{1}{2}\delta \bar{Z}%
_{ii}^{R}+\frac{1}{2}\delta Z_{ii}^{L}\right) m_{i}-\delta m_{i}\,,  \notag
\\
d &=&-\Sigma _{ii}^{R}\left( p^{2}\right) -\left( 1+\frac{1}{2}\delta \bar{Z}%
_{ii}^{L}+\frac{1}{2}\delta Z_{ii}^{R}\right) m_{i}-\delta m_{i}\,.
\label{abcd}
\end{eqnarray}
In the limit $p^{2}\rightarrow m_{i}^{2}$ the chiral structures in the
numerator has to cancel ($a\rightarrow b$ and $c\rightarrow d$), this
requirement leads to
\begin{eqnarray}
\delta \bar{Z}_{ii}^{R}-\delta \bar{Z}_{ii}^{L} &=&\Sigma _{ii}^{\gamma
R}\left( m_{i}^{2}\right) -\Sigma _{ii}^{\gamma L}\left( m_{i}^{2}\right) +%
\frac{\Sigma _{ii}^{R}\left( m_{i}^{2}\right) -\Sigma _{ii}^{L}\left(
m_{i}^{2}\right) }{m_{i}}\,,  \notag \\
\delta Z_{ii}^{R}-\delta Z_{ii}^{L} &=&\Sigma _{ii}^{\gamma R}\left(
m_{i}^{2}\right) -\Sigma _{ii}^{\gamma L}\left( m_{i}^{2}\right) -\frac{%
\Sigma _{ii}^{R}\left( m_{i}^{2}\right) -\Sigma _{ii}^{L}\left(
m_{i}^{2}\right) }{m_{i}}\,.  \label{minus}
\end{eqnarray}
After this, we impose the inverse propagator to have a zero in its real part
as $p^{2}\rightarrow m_{i}^{2}$
\begin{equation}
\lim_{p^{2}\rightarrow m_{i}^{2}}Re\left( p^{2}b-cda^{-1}\right) =0\,,
\end{equation}
from where $\delta m_{i}$ is obtained
\begin{equation}
\delta m_{i}=-\frac{1}{2}Re\left\{ m_{i}\Sigma _{ii}^{\gamma L}\left(
m_{i}^{2}\right) +m_{i}\Sigma _{ii}^{\gamma R}+\Sigma _{ii}^{L}\left(
m_{i}^{2}\right) +\Sigma _{ii}^{R}\left( m_{i}^{2}\right) \right\} \,.
\label{deltam}
\end{equation}
This condition defines a mass and a width that agrees at the one-loop level
with the ones given in \cite{Sirlin:1991fd}, \cite{Sirlin:1998dz}, \cite
{Grassi:2001dz} and \cite{Grassi:2001bz}. Mass and width are defined as the
real an imaginary parts of the propagator pole in the complex plane
respectively. Note also that from Eqs. (\ref{abcd}) (\ref{minus}) and (\ref
{deltam}) we have
\begin{equation}
\lim_{p^{2}\rightarrow m_{i}^{2}}\left( -ca^{-1}\right) =m_{i}+\frac{i}{2}%
Im\left( \Sigma _{ii}^{\gamma R}\left( m_{i}^{2}\right) m_{i}+\Sigma
_{ii}^{\gamma L}\left( m_{i}^{2}\right) m_{i}+\Sigma _{ii}^{R}\left(
m_{i}^{2}\right) +\Sigma _{ii}^{L}\left( m_{i}^{2}\right) \right) \,,
\end{equation}
and therefore
\begin{equation*}
\lim_{p^{2}\rightarrow m_{i}^{2}}\frac{\not{p}\left( aL+bR\right) -dL-cR}{%
p^{2}ab-cd}=\frac{\not{p}+m_{i}-i\Gamma /2}{im_{i}\Gamma }\,,
\end{equation*}
where the width is defined as
\begin{equation*}
\Gamma \equiv -Im\left( \Sigma _{ii}^{\gamma R}\left( m_{i}^{2}\right)
m_{i}+\Sigma _{ii}^{\gamma L}\left( m_{i}^{2}\right) m_{i}+\Sigma
_{ii}^{R}\left( m_{i}^{2}\right) +\Sigma _{ii}^{L}\left( m_{i}^{2}\right)
\right) \,.
\end{equation*}
This quantity is ultraviolet finite. In order to find the residue in the
complex plane we expand the propagator around the physical mass obtaining
for $p^{2}\sim m_{i}^{2}$
\begin{equation}
S_{ij}\left( p\right) =\frac{i\left[ \not{p}+m_{i}-i\Gamma /2+\mathcal{O}%
\left( p^{2}-m_{i}^{2}\right) \right] }{im_{i}\Gamma +\left(
p^{2}-m_{i}^{2}\right) a^{-1}\left[ ab+m_{i}^{2}\left( a^{\prime
}b+ab^{\prime }\right) -\left( c^{\prime }d+cd^{\prime }\right) \right] }+%
\mathcal{O}\left( \left( p^{2}-m_{i}^{2}\right) ^{2}\right) \,,
\end{equation}
where $a=b$ and $c=d$ are evaluated at $p^{2}=m_{i}^{2}$. Hereafter primed
quantities denote derivatives with respect to $p^{2}$. $\mathcal{O}\left(
\left( p^{2}-m_{i}^{2}\right) ^{n}\right) $ stands for non-essential
corrections of order $(p^{2}-m_{i}^{2})^{n}$. Note that the $\mathcal{O}%
\left( p^{2}-m_{i}^{2}\right) $ corrections in the numerator do not mix with
the ones of the same order in the denominator since the first ones are of
order $\Gamma ^{-1}$ and the second ones are of order $\Gamma ^{-2}.$ Taking
into account these comments the unit residue condition amounts to requiring
\begin{equation}
1=\frac{a+b}{2}+m_{i}^{2}\left( a^{\prime }+b^{\prime }\right) +\left(
m_{i}-i\Gamma /2\right) \left( c^{\prime }+d^{\prime }\right) \,,
\end{equation}
from where
\begin{eqnarray}
\frac{1}{2}\left( \delta \bar{Z}_{ii}^{L}+\delta \bar{Z}_{ii}^{R}\right)
&=&\Sigma _{ii}^{\gamma L}\left( m_{i}^{2}\right) +\Sigma _{ii}^{\gamma
R}\left( m_{i}^{2}\right) -\frac{1}{2}\left( \frac{{}}{{}}\delta
Z_{ii}^{L}+\delta Z_{ii}^{R}\frac{{}}{{}}\right) +2m_{i}^{2}\left( \Sigma
_{ii}^{\gamma L\prime }\left( m_{i}^{2}\right) +\Sigma _{ii}^{\gamma R\prime
}\left( m_{i}^{2}\right) \right)  \notag \\
&&+2m_{i}\left( \frac{{}}{{}}\Sigma _{ii}^{L\prime }\left( m_{i}^{2}\right)
+\Sigma _{ii}^{R\prime }\left( m_{i}^{2}\right) \frac{{}}{{}}\right) \,.
\label{residue}
\end{eqnarray}
We have already required all the necessary conditions to fix the correct
properties of the on-shell propagator but still there is some freedom left
in the definition of the diagonal $Z$'s. This freedom can be expressed in
terms of a set of finite coefficients $\alpha _{i}$ given by
\begin{equation*}
\frac{1}{2}\left( \delta Z_{ii}^{L}+\delta Z_{ii}^{R}\right) =\frac{1}{2}%
\left( \delta \bar{Z}_{ii}^{L}+\delta \bar{Z}_{ii}^{R}\right) +\alpha _{i}\,.
\end{equation*}
Bearing in mind that ambiguity and using Eqs. (\ref{minus}) and (\ref
{residue}) we obtain
\begin{eqnarray}
\delta \bar{Z}_{ii}^{L} &=&\Sigma _{ii}^{\gamma L}\left( m_{i}^{2}\right) -X-%
\frac{\alpha _{i}}{2}+D\,,  \notag \\
\delta \bar{Z}_{ii}^{R} &=&\Sigma _{ii}^{\gamma R}\left( m_{i}^{2}\right) +X-%
\frac{\alpha _{i}}{2}+D\,,  \notag \\
\delta Z_{ii}^{L} &=&\Sigma _{ii}^{\gamma L}\left( m_{i}^{2}\right) +X+\frac{%
\alpha _{i}}{2}+D\,,  \notag \\
\delta Z_{ii}^{R} &=&\Sigma _{ii}^{\gamma R}\left( m_{i}^{2}\right) -X+\frac{%
\alpha _{i}}{2}+D\,,  \label{zdiag}
\end{eqnarray}
where
\begin{eqnarray*}
X &=&\frac{1}{2}\frac{\Sigma _{ii}^{R}\left( m_{i}^{2}\right) -\Sigma
_{ii}^{L}\left( m_{i}^{2}\right) }{m_{i}}\,, \\
D &=&m_{i}^{2}\left( \Sigma _{ii}^{\gamma L\prime }\left( m_{i}^{2}\right)
+\Sigma _{ii}^{\gamma R\prime }\left( m_{i}^{2}\right) \right) +m_{i}\left(
\frac{{}}{{}}\Sigma _{ii}^{L\prime }\left( m_{i}^{2}\right) +\Sigma
_{ii}^{R\prime }\left( m_{i}^{2}\right) \frac{{}}{{}}\right) \,.
\end{eqnarray*}
Note that since $X=0$ at the one-loop level and choosing $\alpha _{i}=0$ we
obtain $\delta \bar{Z}_{ii}^{L}=\delta Z_{ii}^{L}$ and $\delta \bar{Z}%
_{ii}^{R}=\delta Z_{ii}^{R}.$ However we have the freedom to choose $\alpha
_{i}\neq 0$. This does not affect the mass terms or neutral current
couplings, but changes the charged coupling currents by multiplying the CKM
matrix $K$ by diagonal matrices. These redefinitions do not change the
physical observables provided the $\alpha _{i}$ are pure imaginary numbers.
This ambiguity corresponds in perturbation theory to the well know freedom
in phase redefinitions of the CKM matrix. Except for this last freedom, the
on-shell conditions determine one unique solution, the one presented here,
with $\bar{Z}^{\frac{1}{2}}\neq \gamma ^{0}Z^{\frac{1}{2}\dagger }\gamma
^{0} $.

\section{W$^{+}$ and top decay}

\renewcommand{\theequation}{\arabic{section}.\arabic{equation}} %
\setcounter{equation}{0}

\label{wandtdecay}Let us now apply the above mechanism to $W^{+}$ and top
decay. We write
\begin{eqnarray}
W^{+}\left( q\right) &\rightarrow &f_{i}\left( p_{1}\right) \bar{f}%
_{j}\left( p_{2}\right) \,,  \label{wdecay} \\
f_{i}\left( p_{1}\right) &\rightarrow &W^{+}\left( q\right) f_{j}\left(
p_{2}\right) \,,  \label{tdecay}
\end{eqnarray}
where $f$ indicates particle and $\bar{f}$ anti-particle. The Latin indices
are reserved for family indices. Leptonic and quark channels can be
considered with the same notation, and confusion should not arise. For the
process (\ref{wdecay}) there are at next-to-leading order two different type
of Lorentz structures
\begin{eqnarray}
M_{L}^{\left( 1\right) } &=&\bar{u}_{i}\left( p_{1}\right) \not{\varepsilon}%
\left( q\right) Lv_{j}\left( p_{2}\right) \,,\qquad \left( L\leftrightarrow
R\right) \,,  \notag \\
M_{L}^{\left( 2\right) } &=&\bar{u}_{i}\left( p_{1}\right) Lv_{j}\left(
p_{2}\right) p_{1}\cdot \varepsilon \left( q\right) \,,\qquad \left(
L\leftrightarrow R\right) \,,  \label{wdtree}
\end{eqnarray}
where $\varepsilon $ stands for the vector polarisation of the $W^{+}$.
Equivalently for the process (\ref{tdecay}) we shall use
\begin{eqnarray}
M_{L}^{\left( 1\right) } &=&\bar{u}_{j}\left( p_{2}\right) \not{\varepsilon}%
^{\ast }\left( q\right) Lu_{i}\left( p_{1}\right) \,,\qquad \left(
L\leftrightarrow R\right) \,,  \notag \\
M_{L}^{\left( 2\right) } &=&\bar{u}_{j}\left( p_{2}\right) Lu_{i}\left(
p_{1}\right) p_{1}\cdot \varepsilon ^{\ast }\left( q\right) \,,\qquad \left(
L\leftrightarrow R\right) \,.  \label{tdtree}
\end{eqnarray}
The transition amplitude at tree level for the processes (\ref{wdecay}) and (%
\ref{tdecay}) is given by
\begin{equation*}
\mathcal{M}_{0}=-\frac{eK_{ij}}{2s_{W}}M_{L}^{\left( 1\right) }\,,
\end{equation*}
where Eq. (\ref{wdtree}) is used for $M_{L}^{\left( 1\right) }$ in $W^{+}$
decay and Eq. (\ref{tdtree}) instead for $M_{L}^{\left( 1\right) }$ in $t$
decay. The one-loop corrected transition amplitude can be written as
\begin{eqnarray}
\mathcal{M}_{1} &=&-\frac{e}{2s_{W}}M_{L}^{\left( 1\right) }\left[
K_{ij}\left( 1+\frac{\delta e}{e}-\frac{\delta s_{W}}{s_{W}}+\frac{1}{2}%
\delta Z_{W}\right) +\delta K_{ij}+\frac{1}{2}\sum_{r}\left( \delta \bar{Z}%
_{ir}^{Lu}K_{rj}+K_{ir}\delta Z_{rj}^{Ld}\right) \right]  \notag \\
&&-\frac{e}{2s_{W}}\left( \delta F_{L}^{\left( 1\right) }M_{L}^{\left(
1\right) }+M_{L}^{\left( 2\right) }\delta F_{L}^{\left( 2\right)
}+M_{R}^{\left( 1\right) }\delta F_{R}^{\left( 1\right) }+M_{R}^{\left(
2\right) }\delta F_{R}^{\left( 2\right) }\right) \,.  \label{vertex}
\end{eqnarray}
In this expression $\delta F_{L,R}^{\left( 1,2\right) }$ are the electroweak
form factors coming from one-loop vertex diagrams. The renormalisation
constants are given by
\begin{eqnarray*}
\frac{\delta e}{e} &=&-\frac{1}{2}\left[ \left( \delta Z_{2}^{A}-\delta
Z_{1}^{A}\right) +\delta Z_{2}^{A}\right] =-\frac{s_{W}}{c_{W}M_{Z}^{2}}\Pi
^{ZA}\left( 0\right) +\frac{1}{2}\frac{\partial \Pi ^{AA}}{\partial k^{2}}%
\left( 0\right) \,, \\
\frac{\delta s_{W}}{s_{W}} &=&-\frac{c_{W}^{2}}{2s_{W}^{2}}\left( \frac{%
\delta M_{W}^{2}}{M_{W}^{2}}-\frac{\delta M_{Z}^{2}}{M_{Z}^{2}}\right) =-%
\frac{c_{W}^{2}}{2s_{W}^{2}}Re\left( \frac{\Pi ^{WW}\left( M_{W}^{2}\right)
}{M_{W}^{2}}-\frac{\Pi ^{ZZ}\left( M_{Z}^{2}\right) }{M_{Z}^{2}}\right) \,,
\\
\delta Z_{W} &=&-\frac{\partial \Pi ^{WW}}{\partial k^{2}}\left(
M_{W}^{2}\right) \,,
\end{eqnarray*}
and the fermionic wfr. constants are depicted in Eqs. (\ref{zin}), (\ref
{zout}) and (\ref{zdiag}) where the indices $u$ or $d$ must be restored in
the masses. The index $A$ refers to the photon field.

As for the $\delta K_{ij}$ renormalisation constants, a SU(2) Ward identity
\cite{Grassi} fixes these counter terms to be
\begin{equation}
\delta K_{jk}=\frac{1}{4}\left[ \left( \delta \hat{Z}^{uL}-\delta \hat{Z}%
^{uL\dagger }\right) K-K\left( \delta \hat{Z}^{dL}-\delta \hat{Z}^{dL\dagger
}\right) \right] _{jk}\,,  \label{deltaK}
\end{equation}
where $\hat{Z}$ means that the wfr. constants appearing in the above
expression are not necessarily the same ones used to renormalise and
guarantee the proper on-shell residue for the external legs as already has
been emphasised. One may, for instance, use minimal subtraction $Z$'s for
the former.

We know \cite{Marciano} that the combination $\frac{\delta e}{e}-\frac{%
\delta s_{W}}{s_{W}}$ is gauge parameter independent. All the other vertex
functions and renormalisation constants are gauge dependent. For the reasons
stated in the introduction we want the amplitude (\ref{vertex}) to be
exactly gauge independent ---not just its modulus--- so the gauge dependence
must cancel between all the remaining terms.

In the next section we shall make use of the Nielsen identities \cite
{Gambino,Nielsen,Piguet,Sibold} to determine that three of the form factors
appearing in the vertex (\ref{vertex}) are by themselves gauge independent,
namely
\begin{equation*}
\partial _{\xi }\delta F_{L}^{\left( 2\right) }=\partial _{\xi }\delta
F_{R}^{\left( 1\right) }=\partial _{\xi }\delta F_{R}^{\left( 2\right) }=0\,.
\end{equation*}
$\xi $ is the gauge-fixing parameter. We shall also see that the gauge
dependence in the remaining form factor $\delta F_{L}^{\left( 1\right) }$
cancels exactly with the one contained in $\delta Z_{W}$ and in $\delta Z$
and $\delta \bar{Z}$. Therefore to guarantee a gauge-fixing parameter
independent amplitude $\delta K$ must be gauge independent as well.

The difficulties related to a proper definition of $\delta K$ were first
pointed out in \cite{Grassi,Gambino}, where it was realized that using the
on-shell $Z$'s of \cite{DennerSack} in Eq.~(\ref{deltaK}) led to a gauge
dependent $K$ and amplitude. They suggested a modification of the on-shell
scheme based on a subtraction at $p^{2}=0$ for all flavours that ensured
gauge independence. We want to stress that the choice for $\delta K$ is not
unique and different choices may differ by gauge independent finite parts
\cite{Kniehl}. Note that the gauge independence of $\delta K$ is in
contradistinction with the conclusions of \cite{Barroso} and in addition
these authors have a non-unitary bare CKM matrix which does not respect the
Ward identity.

As we shall see, if instead of using our prescription for $\delta Z$ and $%
\delta \bar{Z}$ one makes use of the wfr. constants of \cite{Denner} to
renormalise the external fermion legs, it turns out that the gauge
cancellation dictated by the Nielsen identities does not actually take place
in the amplitude. The culprits are of course the absorptive parts. These
absorptive parts of the self-energies are absent in \cite{Denner} due to the
use of the $\widetilde{Re}$ prescription, which throws them away. Notice,
though, that the vertex contribution has gauge dependent absorptive parts
(calculated in the next section) and they remain in the final result.

One might think of absorbing these additional terms in the counter term for $%
\delta K$. This does not work. Indeed one can see from explicit calculations
that wfr. constants decompose as
\begin{equation}
\delta Z^{Lu}=A^{uL}+iB^{uL}\,,\qquad \delta \bar{Z}^{Lu}=A^{uL\dagger
}+iB^{uL\dagger }\,,\qquad \left( L\leftrightarrow R\text{, }%
u\leftrightarrow d\right) \,,  \label{zdecomp}
\end{equation}
where the matrices $A$'s or $B$'s contain the dispersive and absorptive
parts of the self-energies, respectively. Moreover if one substitutes back
Eq. (\ref{zdecomp}) into Eq. (\ref{vertex}) one immediately sees that a
necessary requirement allowing the $A^{u}$ and $A^{d}$ (respectively $B^{u}$ and $%
B^{d}$) contribution to be absorbed into a CKM matrix counter term of the
form given in Eq. (\ref{deltaK}) is that $A^{u}$ and $A^{d}$ (respectively $B^{u}$
and $B^{d}$) were anti-hermitian (respectively hermitian) matrices. By direct
inspection one can conclude that all $A$'s or $B$'s are neither hermitian
nor anti-hermitian matrices and therefore any of such redefinitions are
impossible unless one is willing to give up the unitarity of the bare $K$. A
problem somewhat similar to that was encountered in \cite{Barroso} (but
different, they did not consider absorptive parts at all, the inconsistency
showed up already with the dispersive parts of the on-shell scheme of \cite
{DennerSack}).

It turns out that in the SM these gauge dependent absorptive parts, leading
to a gauge dependent amplitude if they are dropped, do actually cancel, at
least at the one-loop level, in the modulus of the $S$-matrix element. Thus
at this level the use of $\widetilde{Re}$ is irrelevant. It is also shown in
section \ref{absorptive} that gauge independent absorptive parts do survive
even in the modulus of the amplitude for top or anti-top decay (and only in
these cases). Therefore we have to conclude that the difference between
using $\widetilde{Re}$, as advocated in \cite{Denner}, or not, as we do, is
not just a semantic one. As we have seen such difference cannot be
attributed to a finite renormalisation of $K$, provided the bare $K$ remains
unitary as required by the Ward identity (\ref{deltaK}).

\section{Nielsen Identities}

\renewcommand{\theequation}{\arabic{section}.\arabic{equation}} %
\setcounter{equation}{0}

\label{nielsen}

In this section we derive in detail the gauge dependence of the vertex
three-point function. It is therefore rather technical and it can be omitted
by readers just interested in the physical conclusions. In order to have
control on gauge dependence, a useful tool is provided by the so called
Nielsen identities \cite{Nielsen}. For such purpose besides the
``classical'' Lagrangian $\mathcal{L}_{\mathrm{SM}}$ we have to take into
account the gauge fixing term $\mathcal{L}_{\mathrm{GF}}$, the Fadeev-Popov
term $\mathcal{L}_{\mathrm{FP}}$ and source terms. Such source terms are the
ones given by BRST variations of matter ($\bar{\eta}^{u},\eta ^{u},\ldots $)
and gauge fields together with Goldstone and ghost fields (not including
anti-ghosts). We refer the reader to \cite{Hollik}, \cite{Gambino} for
notation and further explanations. We also include source terms ($\chi $)
for the composite operators whose BRST variation generate $\mathcal{L}_{%
\mathrm{GF}}+\mathcal{L}_{\mathrm{FP}}.$ Schematically
\begin{eqnarray*}
\mathcal{L} &=&\mathcal{L}_{\mathrm{SM}}+\mathcal{L}_{\mathrm{GF}}+\mathcal{L%
}_{\mathrm{FP}}-\frac{1}{2\xi }\chi \left( \frac{{}}{{}}\left( \partial
^{\mu }W_{\mu }^{-}-i\xi M_{W}G^{-}\right) \bar{c}^{+}+\left( \partial ^{\mu
}W_{\mu }^{+}-i\xi M_{W}G^{+}\right) \bar{c}^{-}\frac{{}}{{}}\right)  \\
&&+\frac{ig}{\sqrt{2}}\bar{\eta}_{i}^{u}K_{ir}Ld_{r}-\frac{ig}{\sqrt{2}}\bar{%
c}^{+}\bar{d}_{r}K_{rj}^{\dagger }R\eta _{j}^{u}+\bar{s}_{i}^{u}u_{i}+\bar{u}%
_{j}s_{j}^{u}+\bar{s}_{i}^{d}d_{i}+\bar{d}_{j}s_{j}^{d}+\ldots \,,
\end{eqnarray*}
where the ellipsis stands for the remaining source terms. The effective
action, $\Gamma $, is introduced in the standard manner
\begin{equation}
\Gamma \left[ \chi ,\bar{\eta}^{u},\eta ^{u},\bar{u}^{cl},u^{cl},\ldots %
\right] =W\left[ \chi ,\bar{\eta}^{u},\eta ^{u},\bar{s}^{u},s^{u},\ldots %
\right] -\left( \bar{s}_{i}^{u}u_{i}^{cl}+\bar{u}_{j}^{cl}s_{j}^{u}+\bar{s}%
_{i}^{d}d_{i}^{cl}+\bar{d}_{j}^{cl}s_{j}^{d}+\ldots \right) \,,
\label{gamma}
\end{equation}
with
\begin{equation}
e^{iW}=Z\left[ \chi ,\bar{\eta}^{u},\eta ^{u},\bar{s}^{u},s^{u},\ldots %
\right] \equiv \int D\Phi \exp \left( i\mathcal{L}\right) \,.  \label{omega}
\end{equation}
From the above expressions and using BRST transformations we can extract the
Nielsen identities for the three-point functions (see \cite{Nielsen} for
details)
\begin{eqnarray}
\partial _{\xi }\Gamma _{W_{\mu }^{+}\bar{u}_{i}d_{j}} &=&-\Gamma _{\chi
W_{\mu }^{+}\gamma _{W_{\alpha }}^{-}}\Gamma _{W_{\alpha }^{+}\bar{u}%
_{i}d_{j}}-\Gamma _{\chi \bar{u}_{i}\eta _{r}^{u}}\Gamma _{W_{\mu }^{+}%
\bar{u}_{r}d_{j}}-\Gamma _{W_{\mu }^{+}\bar{u}_{i}d_{r}}\Gamma _{\bar{\eta}%
_{r}^{d}d_{j}\chi }-\Gamma _{\chi W_{\mu }^{+}\gamma _{G_{\alpha
}}^{-}}\Gamma _{G_{\alpha }^{+}\bar{u}_{i}d_{j}}  \notag \\
&&-\Gamma _{\chi \gamma _{G_{\alpha }}^{+}\bar{u}_{i}d_{j}}\Gamma
_{G_{\alpha }^{-}W_{\mu }^{+}}-\Gamma _{\chi \gamma _{W_{\alpha }}^{+}\bar{u}%
_{i}d_{j}}\Gamma _{W_{\alpha }^{-}W_{\mu }^{+}}-\Gamma _{\chi W_{\mu }^{+}%
\bar{u}_{i}\eta _{r}^{d}}\Gamma _{\bar{d}_{r}d_{j}}-\Gamma _{\bar{u}%
_{i}u_{r}}\Gamma _{\chi W_{\mu }^{+}\bar{\eta}_{r}^{u}d_{j}}\,,
\label{nielsen00}
\end{eqnarray}
where we have omitted the momentum dependence and defined
\begin{equation*}
\Gamma _{\chi \bar{u}_{i}\eta _{j}^{u}}\equiv \frac{\vec{\delta}}{\delta
\chi }\frac{\vec{\delta}}{\delta \bar{u}_{i}^{cl}\left( p\right) }\frac{%
\delta }{\delta \eta _{j}^{u}\left( p\right) }\Gamma \,,\quad \Gamma _{\bar{%
\eta}_{i}^{u}u_{j}\chi }\equiv \frac{\delta }{\delta \bar{\eta}%
_{i}^{u}\left( p\right) }\frac{\vec{\delta}}{\delta u_{j}^{cl}\left(
p\right) }\frac{\vec{\delta}}{\delta \chi }\Gamma \,.
\end{equation*}
In the rest of this section we shall evaluate the on-shell contributions to
Eq.~(\ref{nielsen00}). Analogously we can also derive Nielsen identities for
two-point functions
\begin{eqnarray}
\partial _{\xi }\Gamma _{W_{\mu }^{+}W_{\beta }^{-}}^{\left( 1\right) }
&=&-2\left( \Gamma _{\chi W_{\mu }^{+}\gamma _{W_{\alpha }}^{-}}^{\left(
1\right) }\Gamma _{W_{\alpha }^{+}W_{\beta }^{-}}+\Gamma _{\chi W_{\mu
}^{+}\gamma _{G_{\alpha }}^{-}}^{\left( 1\right) }\Gamma _{G_{\alpha
}^{+}W_{\beta }^{-}}\right) \,,  \label{bose1} \\
\partial _{\xi }\Gamma _{W_{\mu }^{+}G_{\beta }^{-}}^{\left( 1\right) }
&=&-2\left( \Gamma _{\chi W_{\mu }^{+}\gamma _{W_{\alpha }}^{-}}^{\left(
1\right) }\Gamma _{W_{\alpha }^{+}G_{\beta }^{-}}+\Gamma _{\chi W_{\mu
}^{+}\gamma _{G_{\alpha }}^{-}}^{\left( 1\right) }\Gamma _{G_{\alpha
}^{+}G_{\beta }^{-}}\right) \,.  \label{bose2}
\end{eqnarray}
On-shell these reduce to
\begin{equation}
\Gamma _{\chi W^{+}\gamma _{W}^{-}}^{T\left( 1\right) }\left(
M_{W}^{2}\right) =-\frac{1}{2}\partial _{\xi }\left. \frac{\partial \Gamma
_{W^{+}W^{-}}^{T\left( 1\right) }}{\partial q^{2}}\left( q^{2}\right)
\right| _{q^{2}=M_{W}^{2}}=\frac{1}{2}\partial _{\xi }\delta Z_{W}\,,\quad
\Gamma _{\chi W^{+}\gamma _{G}^{-}}^{T\left( 1\right) }\left( q\right) =0\,,
\label{nielsenp}
\end{equation}
where the superscript $T$ refers to the transverse part and the superscript $%
(1)$ makes reference to the one-loop order correction.

Using these two sets of results and restricting Eq.~(\ref{nielsen00}) to the
1PI function appropriate for (on-shell) top-decay
\begin{eqnarray}
&&\bar{u}_{u}\left( p_{i}\right) \epsilon ^{\mu }\left( q\right) \partial
_{\xi }\Gamma _{W_{\mu }^{+}\bar{u}_{i}d_{j}}^{\left( 1\right) }v_{d}\left(
-p_{j}\right)  
\notag \\
&=&\frac{g}{\sqrt{2}}\bar{u}_{u}\left( p_{i}\right) \left\{ \Gamma _{\chi
\bar{u}_{i}\eta _{r}^{u}}K_{rj}\not{\epsilon}L+K_{ir}\not{\epsilon}L\Gamma _{%
\bar{\eta}_{r}^{d}d_{j}\chi }+\frac{1}{2}\partial _{\xi }\delta Z_{W}K_{ij}%
\not{\epsilon}L\right\} v_{d}\left( -p_{j}\right) \,.  \label{nielsen0}
\end{eqnarray}

At the one-loop level we also have the Nielsen identity
\begin{equation}
\partial _{\xi }\Sigma _{ij}^{u}\left( p\right) =\Gamma _{\chi \bar{u}%
_{i}\eta _{j}^{u}}^{\left( 1\right) }\left( p\right) \left( \not{p}%
-m_{j}^{u}\right) +\left( \not{p}-m_{i}^{u}\right) \Gamma _{\bar{\eta}%
_{i}^{u}u_{j}\chi }^{\left( 1\right) }\left( p\right) \,,  \label{nielsen2}
\end{equation}
which is the fermionic counterpart of Eqs. (\ref{bose1}) and (\ref{bose2}).
Similar relation holds interchanging $u\leftrightarrow d$. With the use of
Eq.~(\ref{nielsen2}) and an analogous decomposition to Eq. (\ref{decomp})
for $\Gamma $,
\begin{eqnarray}
\Gamma _{\chi \bar{u}_{i}\eta _{j}^{u}}^{\left( 1\right) }\left( p\right) &=&%
\not{p}\left( \Gamma _{\chi \bar{u}_{i}\eta _{j}^{u}}^{\gamma R\left(
1\right) }\left( p^{2}\right) R+\Gamma _{\chi \bar{u}_{i}\eta
_{j}^{u}}^{\gamma L\left( 1\right) }\left( p^{2}\right) L\right) +\Gamma
_{\chi \bar{u}_{i}\eta _{j}^{u}}^{R\left( 1\right) }\left( p^{2}\right)
R+\Gamma _{\chi \bar{u}_{i}\eta _{j}^{u}}^{L\left( 1\right) }\left(
p^{2}\right) L\,,  \notag \\
\Gamma _{\bar{\eta}_{i}^{u}u_{j}\chi }^{\left( 1\right) }\left( p\right) &=&%
\not{p}\left( \Gamma _{\bar{\eta}_{i}^{u}u_{j}\chi }^{\gamma R\left(
1\right) }\left( p^{2}\right) R+\Gamma _{\bar{\eta}_{i}^{u}u_{j}\chi
}^{\gamma L\left( 1\right) }\left( p^{2}\right) L\right) +\Gamma _{\bar{\eta}%
_{i}^{u}u_{j}\chi }^{R\left( 1\right) }\left( p^{2}\right) R+\Gamma _{\bar{%
\eta}_{i}^{u}u_{j}\chi }^{L\left( 1\right) }\left( p^{2}\right) L\,,
\label{decomp2}
\end{eqnarray}
we obtain after equating Dirac structures
\begin{eqnarray}
\partial _{\xi }\Sigma _{ij}^{u\gamma R}\left( p^{2}\right) &=&\Gamma _{\chi
\bar{u}_{i}\eta _{j}^{u}}^{L\left( 1\right) }\left( p^{2}\right)
-m_{j}\Gamma _{\chi \bar{u}_{i}\eta _{j}^{u}}^{\gamma R\left( 1\right)
}\left( p^{2}\right) +\Gamma _{\bar{\eta}_{i}^{u}u_{j}\chi }^{R\left(
1\right) }\left( p^{2}\right) -m_{i}\Gamma _{\bar{\eta}_{i}^{u}u_{j}\chi
}^{\gamma R\left( 1\right) }\left( p^{2}\right) \,,  \notag \\
\partial _{\xi }\Sigma _{ij}^{uR}\left( p^{2}\right) &=&p^{2}\Gamma _{\chi
\bar{u}_{i}\eta _{j}^{u}}^{\gamma L\left( 1\right) }\left( p^{2}\right)
-m_{j}\Gamma _{\chi \bar{u}_{i}\eta _{j}^{u}}^{R\left( 1\right) }\left(
p^{2}\right) +p^{2}\Gamma _{\bar{\eta}_{i}^{u}u_{j}\chi }^{\gamma R\left(
1\right) }\left( p^{2}\right) -m_{i}\Gamma _{\bar{\eta}_{i}^{u}u_{j}\chi
}^{R\left( 1\right) }\left( p^{2}\right) \,,  \label{rels}
\end{eqnarray}
and analogous expressions exchanging $L\leftrightarrow R$ and $%
u\leftrightarrow d$. Moreover from Eqs. (\ref{nielsen0}) and (\ref{decomp2})
we obtain
\begin{eqnarray}
&&\hspace{-0.8cm}\bar{u}_{u}\left( p_{i}\right) \epsilon ^{\mu }\left(
q\right) \partial _{\xi }\Gamma _{W_{\mu }^{+}\bar{u}_{i}d_{j}}^{\left(
1\right) }v_{d}\left( -p_{j}\right) =\frac{g}{\sqrt{2}}\left\{ \frac{{}}{{}}%
\bar{u}_{u}\left( p_{i}\right) \left( m_{i}^{u}\Gamma _{\chi \bar{u}_{i}\eta
_{r}^{u}}^{\gamma R\left( 1\right) }\left( m_{i}^{u2}\right) +\Gamma _{\chi
\bar{u}_{i}\eta _{r}^{u}}^{R\left( 1\right) }\left( m_{i}^{u2}\right)
\right) K_{rj}\not{\epsilon}Lv_{d}\left( -p_{j}\right) \right. +  \notag \\
&&\hspace{-0.9cm}\bar{u}_{u}\left( p_{i}\right) K_{ir}\not{\epsilon}L\left(
m_{j}^{d}\Gamma _{\bar{\eta}_{r}^{d}d_{j}\chi }^{\gamma R\left( 1\right)
}\left( m_{j}^{d2}\right) +\Gamma _{\bar{\eta}_{r}^{d}d_{j}\chi }^{L\left(
1\right) }\left( m_{j}^{d2}\right) \right) v_{d}\left( -p_{j}\right) +\left.
\frac{1}{2}\partial _{\xi }\delta Z_{W}\bar{u}_{u}\left( p_{i}\right) K_{ij}%
\not{\epsilon}Lv_{d}\left( -p_{j}\right) \frac{{}}{{}}\right\} \,.
\label{final}
\end{eqnarray}
Using Eqs.~(\ref{zin}), (\ref{zout}) and (\ref{rels}) one arrives at
\begin{eqnarray}
m_{j}^{u}\Gamma _{\bar{\eta}_{i}^{u}u_{j}\chi }^{\gamma R\left( 1\right)
}\left( m_{j}^{u2}\right) +\Gamma _{\bar{\eta}_{i}^{u}u_{j}\chi }^{L\left(
1\right) }\left( m_{j}^{u2}\right) &=&\frac{1}{2}\partial _{\xi }\delta
Z_{ij}^{uL}\,,\quad \left( i\neq j\right) \,,  \label{gammalin} \\
m_{i}^{u}\Gamma _{\chi \bar{u}_{i}\eta _{j}^{u}}^{\gamma R\left( 1\right)
}\left( m_{i}^{u2}\right) +\Gamma _{\chi \bar{u}_{i}\eta _{j}^{u}}^{R\left(
1\right) }\left( m_{i}^{u2}\right) &=&\frac{1}{2}\partial _{\xi }\delta \bar{%
Z}_{ij}^{uL}\,,\quad \left( i\neq j\right) \,,  \label{gammalout}
\end{eqnarray}
and once more similar relations hold exchanging $L\leftrightarrow R$ and $%
u\leftrightarrow d$. Notice that absorptive parts are present in the 1PI
Green functions and hence in $\delta Z$ and $\delta \bar{Z}$ too. If we
forget about such absorptive parts we would have pseudo-hermiticity. Namely
\begin{equation*}
\Gamma _{\chi \bar{u}_{i}\eta _{j}^{u}}^{\left( 1\right) }=\gamma ^{0}\Gamma
_{\bar{\eta}_{i}^{u}u_{j}\chi }^{\left( 1\right) \dagger }\gamma ^{0}\,,
\end{equation*}
where $\Gamma _{\bar{\eta}_{i}^{u}u_{j}\chi }^{\dagger }$ means complex
conjugating $\Gamma _{\bar{\eta}_{i}^{u}u_{j}\chi }$ and interchanging \emph{%
both} Dirac and family indices. However the imaginary branch cuts terms
prevent the above relation to hold and then Eq. (\ref{hermiticity}) does not
hold.

At this point one might be tempted to plug expressions (\ref{gammalin}), (%
\ref{gammalout}) in Eq. (\ref{final}). However such relations are obtained
only in the restricted case $i\neq j$. For $i=j$ Eqs. (\ref{rels}) are
insufficient to determine the combinations appearing in the lhs. of Eqs. (%
\ref{gammalin}), (\ref{gammalout}) and further information is required. That
is also necessary even in the actual case where the rhs. of Eqs. (\ref
{gammalin}), (\ref{gammalout}) are not singular at $m_{i}\rightarrow m_{j}$
\cite{Yamada}. In the rest of this section we shall proceed to calculate
such diagonal combinations and as by product we shall also cross-check the
results already obtained for the off-diagonal contributions and in addition
produce some new ones.

By direct computation one generically finds
\begin{eqnarray}
\Gamma _{\chi \bar{u}_{i}\eta _{j}^{u}}^{\left( 1\right) } &=&\left( \frac{{}%
}{{}}\not{p}m_{i}^{u}B_{ij}^{u}\left( p^{2}\right) +C_{ij}^{u}\left(
p^{2}\right) +A_{ij}^{u}\left( p^{2}\right) \frac{{}}{{}}\right) R\,,  \notag
\\
\Gamma _{\bar{\eta}_{i}^{u}u_{j}\chi }^{\left( 1\right) } &=&L\left( \frac{{}%
}{{}}\not{p}B_{ij}^{u}\left( p^{2}\right) m_{j}^{u}+C_{ij}^{u}\left(
p^{2}\right) +A_{ij}^{u}\left( p^{2}\right) \frac{{}}{{}}\right) \,,
\label{integrals}
\end{eqnarray}
and analogous relations interchanging $u\leftrightarrow d.$ The $A$ function
comes from the diagram containing a charged gauge boson propagator and $B$
and $C$ from the diagram containing a charged Goldstone boson propagator.
From Eqs. (\ref{nielsen2}) and (\ref{integrals}) we obtain
\begin{eqnarray}
\partial _{\xi }\Sigma _{ij}^{\gamma R}\left( p^{2}\right)
&=&-2m_{i}B_{ij}\left( p^{2}\right) m_{j}\,,  \notag \\
\partial _{\xi }\Sigma _{ij}^{\gamma L}\left( p^{2}\right) &=&2\left( \frac{%
{}}{{}}A_{ij}\left( p^{2}\right) +C_{ij}\left( p^{2}\right) \frac{{}}{{}}%
\right) \,,  \notag \\
\partial _{\xi }\Sigma _{ij}^{R}\left( p^{2}\right) &=&\left( \frac{{}}{{}}%
p^{2}B_{ij}\left( p^{2}\right) -C_{ij}\left( p^{2}\right) -A_{ij}\left(
p^{2}\right) \frac{{}}{{}}\right) m_{j}\,,  \notag \\
\partial _{\xi }\Sigma _{ij}^{L}\left( p^{2}\right) &=&m_{i}\left( \frac{{}}{%
{}}p^{2}B_{ij}\left( p^{2}\right) -C_{ij}\left( p^{2}\right) -A_{ij}\left(
p^{2}\right) \frac{{}}{{}}\right) \,.  \label{sigs}
\end{eqnarray}
The above system of equations is overdetermined and therefore some
consistency identities between bare self-energies arise, namely
\begin{equation}
\partial _{\xi }\left( \frac{{}}{{}}m_{i}\Sigma _{ij}^{R}\left( p^{2}\right)
-\Sigma _{ij}^{L}\left( p^{2}\right) m_{j}\frac{{}}{{}}\right) =0\,,
\label{const0}
\end{equation}
and
\begin{equation}
\partial _{\xi }\left( p^{2}\Sigma _{ij}^{\gamma R}\left( p^{2}\right)
+\Sigma _{ij}^{\gamma L}\left( p^{2}\right) m_{i}m_{j}+m_{i}\Sigma
_{ij}^{R}\left( p^{2}\right) +\Sigma _{ij}^{L}\left( p^{2}\right)
m_{j}\right) =0\,.  \label{const2}
\end{equation}
These constrains must hold independently of any renormalisation scheme and
we have checked them by direct computation. Actually the former trivially
holds since, at least at the one-loop level in the SM,
\begin{equation}
m_{i}\Sigma _{ij}^{R}\left( p^{2}\right) -\Sigma _{ij}^{L}\left(
p^{2}\right) m_{j}=0\,.  \label{const1}
\end{equation}

Finally, projecting Eq. (\ref{integrals}) over spinors we also have
\begin{eqnarray}
\bar{u}_{u}\left( p_{i}\right) \Gamma _{\chi \bar{u}_{i}\eta
_{j}^{u}}^{\left( 1\right) } &=&\bar{u}_{u}\left( p_{i}\right) \left( \frac{%
{}}{{}}m_{i}^{u2}B_{ij}^{u}\left( m_{i}^{u2}\right) +C_{ij}^{u}\left(
m_{i}^{u2}\right) +A_{ij}^{u}\left( m_{i}^{u2}\right) \frac{{}}{{}}\right)
R\,,  \notag \\
\Gamma _{\bar{\eta}_{i}^{d}d_{j}\chi }^{\left( 1\right) }v_{d}\left(
-p_{j}\right) &=&L\left( \frac{{}}{{}}B_{ij}^{d}\left( m_{j}^{d2}\right)
m_{j}^{d2}+C_{ij}^{d}\left( m_{j}^{d2}\right) +A_{ij}^{d}\left(
m_{j}^{d2}\right) \right) v_{d}\left( -p_{j}\right) \,.  \label{calc0}
\end{eqnarray}
The rhs. of the previous expressions can be evaluated in terms of the wfr.
via the use of Eqs. (\ref{sigs})
\begin{eqnarray}
&\partial _{\xi }&\hspace{-0.3cm}\left( m_{j}^{u}m_{i}^{u}\Sigma
_{ij}^{u\gamma R}\left( p^{2}\right) +p^{2}\Sigma _{ij}^{u\gamma L}\left(
p^{2}\right) +m_{j}^{u}\Sigma _{ij}^{uR}\left( p^{2}\right) +m_{i}^{u}\Sigma
_{ij}^{uL}\left( p^{2}\right) \right) =  \notag \\
&&B_{ij}^{u}\left( p^{2}\right) \left( \frac{{}}{{}}p^{2}\left(
m_{j}^{u2}+m_{i}^{u2}\right) -2m_{j}^{u2}m_{i}^{u2}\frac{{}}{{}}\right)
+\left( 2p^{2}-m_{j}^{u2}-m_{i}^{u2}\right) \left( A_{ij}^{u}\left(
p^{2}\right) +C_{ij}^{u}\left( p^{2}\right) \frac{{}}{{}}\right) \,,
\label{calcu} \\
&\partial _{\xi }&\hspace{-0.3cm}\left( \Sigma _{ij}^{d\gamma R}\left(
p^{2}\right) m_{i}^{d}m_{j}^{d}+\Sigma _{ij}^{d\gamma L}\left( p^{2}\right)
p^{2}+m_{i}^{d}\Sigma _{ij}^{dL}\left( p^{2}\right) +\Sigma _{ij}^{dR}\left(
p^{2}\right) m_{j}^{d}\right) =  \notag \\
&&B_{ij}^{d}\left( p^{2}\right) \left( \frac{{}}{{}}p^{2}\left(
m_{i}^{d2}+m_{j}^{d2}\right) -2m_{i}^{d2}m_{j}^{d2}\right) +\left(
2p^{2}-m_{i}^{d2}-m_{j}^{d2}\right) \left( \frac{{}}{{}}A_{ij}^{d}\left(
p^{2}\right) +C_{ij}^{d}\left( p^{2}\right) \right) \,.  \label{calcd}
\end{eqnarray}
Hence using the off-diagonal wfr. expressions (\ref{zin}), (\ref{zout}) we
re-obtain
\begin{equation}
\bar{u}_{u}\left( p_{i}\right) \frac{1}{2}\partial _{\xi }\delta \bar{Z}%
_{ij}^{uL}R=\bar{u}\left( p_{i}\right) \Gamma _{\chi \bar{u}_{i}\eta
_{j}^{u}}^{\left( 1\right) }\,,\qquad L\frac{1}{2}\partial _{\xi }\delta
Z_{ij}^{dL}v_{d}\left( -p_{j}\right) =\Gamma _{\bar{\eta}_{i}^{d}d_{j}\chi
}^{\left( 1\right) }v_{d}\left( -p_{j}\right) \,.  \label{basicrel}
\end{equation}
For the diagonal wfr. we use Eqs. (\ref{zdiag}) together with (\ref{sigs})
and (\ref{calc0}) obtaining \emph{exactly} the same result as in Eq. (\ref
{basicrel}) with $i=j$ therein. Note however that since in Eq. (\ref{calc0})
we have no derivatives with respect to $p^{2}$ obtaining Eq. (\ref{basicrel}%
) involves a subtle cancellation between the $p^{2}$ derivatives of the bare
self-energies appearing in the definition of the diagonal wfr.

Before proceeding let us make a side remark concerning the regularity
properties of the gauge derivative in Eqs. (\ref{calcu}) and (\ref{calcu})
in the limit $m_{i}\rightarrow m_{j}$. Note that evaluating Eq. (\ref{calcu}%
) at $p^{2}=m_{i}^{u2}$ and Eq. (\ref{calcd}) at $p^{2}=m_{j}^{d2},$ a
global factor $\left( m_{i}^{u2}-m_{j}^{u2}\right) $ appears in the first
equation and $\left( m_{j}^{d2}-m_{i}^{d2}\right) $ in the second one.
Therefore it can be immediately seen that Nielsen identities together with
the information provided by Eq. (\ref{integrals}) assures the regularity of
the gauge derivative for the off-diagonal wfr. constants when $%
m_{i}\rightarrow m_{j}$. Moreover we have seen that such limit is not only
regular but also equal to the expression obtained from the diagonal wfr.
which is not \emph{a priori} obvious \cite{Grassi}, \cite{Yamada}.

\begin{figure}[t]
\begin{center}
\includegraphics[width=\textwidth]{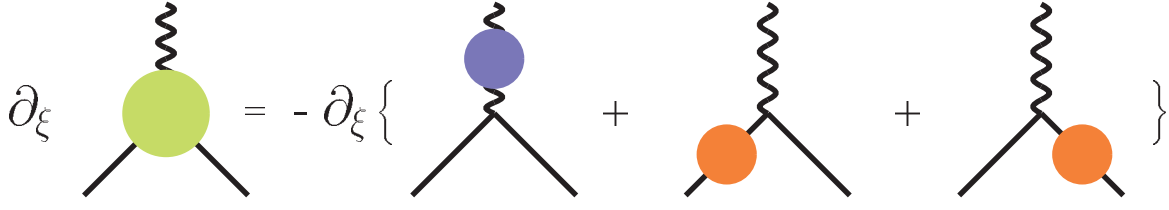}
\end{center}
\caption{Pictorial representation of the on-shell Nielsen identity given by
Eq.(\ref{vertexgauge}). The blobs in the lhs. represent bare one-loop
contributions to the on-shell vertex and the blobs in the rhs. wfr. counter
terms.}
\label{fig1}
\end{figure}

Replacing Eq. (\ref{basicrel}) in Eq. (\ref{nielsen0}) we obtain
\begin{eqnarray}
\partial _{\xi }\left( \bar{u}_{u}\left( p_{i}\right) \epsilon ^{\mu }\left(
q\right) \Gamma _{W_{\mu }^{+}\bar{u}_{i}d_{j}}^{\left( 1\right)
}v_{d}\left( -p_{j}\right) \right) &=&\frac{e}{2s_{W}}M_{L}^{\left( 1\right)
}\partial _{\xi }\left( \delta \bar{Z}_{ir}^{uL}K_{rj}+K_{ir}\delta
Z_{rj}^{dL}+\delta Z_{W}K_{ij}\right)  \notag \\
&=&-\frac{e}{2s_{W}}\partial _{\xi }\left( M_{L}^{\left( 1\right) }\delta
F_{L}^{\left( 1\right) }+M_{L}^{\left( 2\right) }\delta F_{L}^{\left(
2\right) }+M_{R}^{\left( 1\right) }\delta F_{R}^{\left( 1\right)
}+M_{R}^{\left( 2\right) }\delta F_{R}^{\left( 2\right) }\right) \,,  \notag
\\
&&  \label{vertexgauge}
\end{eqnarray}
where Eq.~(\ref{vertex}) and the gauge independence of the electric charge
and Weinberg angle has been used in the last equality. In the previous
expression $M_{L,R}^{\left( i\right) }$ are understood with the physical
momenta $p_{1}$ and $p_{2}$ of Eq. (\ref{wdtree}) replaced by the
diagrammatic momenta $p_{i}$ and $-p_{j}$ respectively. Note that Eq. (\ref
{vertexgauge}) states that the gauge dependence of the on-shell bare
one-loop vertex function cancels out the renormalisation counter terms
appearing in Eq. (\ref{vertex}) (see Fig. \ref{fig1}). This is one of the
crucial results and special care should be taken not to ignore any of the
absorptive parts ---including those in the wfr. constants. As a consequence
\begin{equation*}
\partial _{\xi }\mathcal{M}_{1}=-\frac{e}{2s_{W}}M_{L}^{\left( 1\right)
}\partial _{\xi }\delta K_{ij}\,,
\end{equation*}
and asking for a gauge independent amplitude the counter term for $K_{ij}$
must be separately gauge independent, as originally derived in \cite{Grassi}.

Finally, since each structure $M_{L,R}^{\left( i\right) }$ must cancel
separately we have that the Nielsen identities enforce
\begin{equation*}
\partial _{\xi }\delta F_{L}^{\left( 2\right) }=\partial _{\xi }\delta
F_{R}^{\left( 1\right) }=\partial _{\xi }\delta F_{R}^{\left( 2\right) }=0\,.
\end{equation*}

\section{Absorptive parts}

\label{absorptive} Having determined in the previous section, thanks to an
extensive use of the Nielsen identities, the gauge dependence of the
different quantities appearing in top or $W$ decay in terms of the
self-energies, we shall now proceed to list the absorptive parts of the wfr.
constants, with special attention to their gauge dependence. The aim of this
section is to state the differences between the wfr. constants given in our
scheme and the ones in \cite{Denner}. Recall that at one-loop such
difference reduces to the absorptive ($\widetilde{Im}$) contribution to the $%
\delta Z$'s. In what concerns the gauge dependent part (with $\xi \geq 0$)
the absorptive contribution ($\widetilde{Im}_{\xi }$) in the fermionic $%
\delta Z$'s amounts to

\begin{eqnarray*}
i\widetilde{Im}_{\xi }\left( \delta Z_{ij}^{uL}\right) &=&\sum_{h}\frac{%
iK_{ih}K_{hj}^{\dagger }}{8\pi v^{2}m_{j}^{u2}}\theta \left(
m_{j}^{u}-m_{h}^{d}-\sqrt{\xi }M_{W}\right) \left( m_{j}^{u2}-m_{h}^{d2}-\xi
M_{W}^{2}\right) \\
&&\times \sqrt{\left( \left( m_{j}^{u}-m_{h}^{d}\right) ^{2}-\xi
M_{W}^{2}\right) \left( \left( m_{j}^{u}+m_{h}^{d}\right) ^{2}-\xi
M_{W}^{2}\right) }\,, \\
i\widetilde{Im}_{\xi }\left( \delta \bar{Z}_{ij}^{uL}\right) &=&\sum_{h}%
\frac{iK_{ih}K_{hj}^{\dagger }}{8\pi v^{2}m_{i}^{u2}}\theta \left(
m_{i}^{u}-m_{h}^{d}-\sqrt{\xi }M_{W}\right) \left( m_{i}^{u2}-m_{h}^{d2}-\xi
M_{W}^{2}\right) \\
&&\times \sqrt{\left( \left( m_{i}^{u}-m_{h}^{d}\right) ^{2}-\xi
M_{W}^{2}\right) \left( \left( m_{i}^{u}+m_{h}^{d}\right) ^{2}-\xi
M_{W}^{2}\right) }\,, \\
\widetilde{Im}_{\xi }\left( \delta Z_{ij}^{uR}\right) &=&\widetilde{Im}_{\xi
}\left( \delta \bar{Z}_{ij}^{uR}\right) =0\,,
\end{eqnarray*}
where $\theta $ is the Heaviside function and $v$ is the Higgs vacuum
expectation value. For the down $\delta Z$ we have the same formulae
replacing $u\leftrightarrow d$ and $K\leftrightarrow K^{\dagger }.$ Note
that using these results we can write
\begin{eqnarray}
&&i\partial _{\xi }\widetilde{Im}\left[ \sum_{r}\left( \delta \bar{Z}%
_{ir}^{uL}K_{rj}+K_{ir}\delta Z_{rj}^{dL}\right) +\delta Z_{W}K_{ij}\right]
\notag \\
&=&K_{ij}\partial _{\xi }\left\{ \frac{i}{8\pi v^{2}}\left[ \frac{1}{%
m_{i}^{u2}}\theta \left( m_{i}^{u}-m_{j}^{d}-\sqrt{\xi }M_{W}\right) \left(
m_{i}^{u2}-m_{j}^{d2}-\xi M_{W}^{2}\right) \right. \right.  \notag \\
&&+\left. \frac{1}{m_{j}^{d2}}\theta \left( m_{j}^{d}-m_{i}^{u}-\sqrt{\xi }%
M_{W}\right) \left( m_{j}^{d2}-m_{i}^{u2}-\xi M_{W}^{2}\right) \frac{{}}{{}}%
\right]  \notag \\
&&\times \left. \sqrt{\left( \left( m_{j}^{d}-m_{i}^{u}\right) ^{2}-\xi
M_{W}^{2}\right) \left( \left( m_{j}^{d}+m_{i}^{u}\right) ^{2}-\xi
M_{W}^{2}\right) }+i\widetilde{Im}_{\xi }\left( \delta Z_{W}\right) \frac{{}%
}{{}}\right\} \,.  \label{gauge0}
\end{eqnarray}
In the case $\left| m_{i}^{u}-m_{j}^{d}\right| \leq \sqrt{\xi }M_{W}$ the
above expression reduces to
\begin{equation}
\partial _{\xi }\sum_{r}\widetilde{Im}\left( \delta \bar{Z}%
_{ir}^{uL}K_{rj}+K_{ir}\delta Z_{rj}^{dL}\right) =0\,,  \label{gauge1}
\end{equation}
while for $\left| m_{i}^{u}-m_{j}^{d}\right| \geq \sqrt{\xi }M_{W}$ we have
\begin{eqnarray}
&&\hspace{-0.7cm}i\partial _{\xi }\sum_{r}\widetilde{Im}\left( \delta \bar{Z}%
_{ir}^{uL}K_{rj}+K_{ir}\delta Z_{rj}^{dL}\right) =  \notag \\
&&\hspace{-0.7cm}K_{ij}\partial _{\xi }\left\{ \frac{i}{4\pi v^{2}}\frac{%
\left| m_{i}^{u2}-m_{j}^{d2}\right| -\xi M_{W}^{2}}{m_{i}^{u2}+m_{j}^{d2}+%
\left| m_{i}^{u2}-m_{j}^{d2}\right| }\sqrt{\left( \left(
m_{j}^{d}-m_{i}^{u}\right) ^{2}-\xi M_{W}^{2}\right) \left( \left(
m_{j}^{d}+m_{i}^{u}\right) ^{2}-\xi M_{W}^{2}\right) }\right\} \,.
\label{gauge2}
\end{eqnarray}
Moreover the $\xi $-dependent absorptive contribution to $\delta Z_{W}$ ($%
\widetilde{Im}_{\xi }\left( \delta Z_{W}\right) $) has no dependence in
quark masses since the diagram with a fermion loop is gauge independent.
Because of that we can conclude that the derivative in Eq. (\ref{gauge0})
does not vanish. Defining $\Delta _{ij}$ as the difference between the
vertex observable calculated in our scheme and the same in the scheme using $%
\widetilde{Re}$ we have
\begin{equation*}
\Delta _{ij}\sim \left| K_{ij}\right| ^{2}\mathrm{Re}\left( i\widetilde{Im}%
\delta Z_{W}\right) +\mathrm{Re}\left\{ iK_{ij}^{\ast }\sum_{r}\left[
\widetilde{Im}\left( \delta \bar{Z}_{ir}^{uL}\right) K_{rj}+K_{ir}\widetilde{%
Im}\left( \delta Z_{rj}^{dL}\right) \right] \right\} \,.
\end{equation*}
In the case of $\delta Z_{W}$ one can easily check that $\widetilde{Im}%
\left( \delta Z_{W}\right) =\mathrm{Im}\left( \delta Z_{W}\right) $
obtaining
\begin{equation}
\Delta _{ij}\sim \mathrm{Re}\left\{ iK_{ij}^{\ast }\sum_{r}\left[ \widetilde{%
Im}\left( \delta \bar{Z}_{ir}^{uL}\right) K_{rj}+K_{ir}\widetilde{Im}\left(
\delta Z_{rj}^{dL}\right) \right] \right\} \,.  \label{gauge3}
\end{equation}
Thus from Eqs. (\ref{gauge1}), (\ref{gauge2}) and (\ref{gauge3}) we
immediately obtain
\begin{equation}
\partial _{\xi }\Delta _{ij}\sim \mathrm{Re}\left\{ iK_{ij}^{\ast }\sum_{r}%
\left[ \partial _{\xi }\widetilde{Im}\left( \delta \bar{Z}_{ir}^{uL}\right)
K_{rj}+K_{ir}\partial _{\xi }\widetilde{Im}\left( \delta Z_{rj}^{dL}\right) %
\right] \right\} =0\,.  \label{res1}
\end{equation}
However gauge independent absorptive parts, included if our prescription is
used but not if one uses the one of \cite{Denner} which makes use of the $%
\widetilde{Re}$, do contribute to Eq. (\ref{gauge3}). In order to see that
we can take $\xi =1$ obtaining for the physical values of the masses
\begin{eqnarray}
\widetilde{Im}_{\xi =1}\left( \delta Z_{rj}^{dL}\right) &=&0\,,  \notag \\
\widetilde{Im}_{\xi =1}\left( \delta \bar{Z}_{ir}^{uL}\right) &=&\sum_{h}%
\frac{K_{ih}K_{hr}^{\dagger }}{8\pi v^{2}m_{i}^{u2}}\frac{\theta \left(
m_{i}^{u}-m_{h}^{d}-M_{W}\right) }{m_{i}^{u2}-m_{r}^{u2}}\sqrt{\left(
m_{i}^{u2}-\left( M_{W}-m_{h}^{d}\right) ^{2}\right) \left(
m_{i}^{u2}-\left( M_{W}+m_{h}^{d}\right) ^{2}\right) }  \notag \\
&&\times \left( \frac{1}{2}\left( m_{r}^{u2}+m_{h}^{d2}+2M_{W}^{2}\right)
\left( m_{i}^{u2}+m_{h}^{2d}-M_{W}^{2}\right) -\left(
m_{i}^{u2}+m_{r}^{u2}\right) m_{h}^{d2}\right) \,,  \label{zulb}
\end{eqnarray}
where only the results for $i\neq j$ have been presented. Note that $%
\widetilde{Im}_{\xi =1}\left( \delta \bar{Z}_{ir}^{uL}\right) \neq 0$ only
when $i=3$, that is when the renormalised up-particle is a top. In addition,
since the $m_{r}^{u2}$ dependence in Eq. (\ref{zulb}) does not vanish, CKM
phases do not disappear from Eq. (\ref{gauge3}) and therefore
\begin{equation}
\Delta _{3j}\sim \mathrm{Re}\left\{ iK_{3j}^{\ast }\sum_{r}\left[ \widetilde{%
Im}\left( \delta \bar{Z}_{3r}^{uL}\right) K_{rj}+K_{3r}\widetilde{Im}\left(
\delta Z_{rj}^{dL}\right) \right] \right\} \neq 0\,.  \label{res2}
\end{equation}
Eqs. (\ref{res1}) and (\ref{res2}) show that even though the difference $%
\Delta _{3j}$ is gauge independent, does not actually vanish. There are
genuine gauge independent pieces that contribute not only to the amplitude,
but also to the observable. As discussed these additional pieces cannot be
absorbed by a redefinition of $K_{ij}$. Numerically such gauge independent
corrections amounts roughly to $\Delta _{3j}\simeq 5\times 10^{-3}O_{\mathrm{%
tree}}$ where $O_{\mathrm{tree}}$ is the observable quantity calculated at
leading order.

\section{CP violation and CPT invariance}

\label{cp}In this section we want to show that using wfr. constants that do
not verify a pseudo-hermiticity condition does not lead to any unwanted
pathologies. In particular: (a) No new sources of $CP$ violation appear
besides the ones already present in the SM. (b) The total width of particles
and anti-particles coincide, thus verifying the $CPT$ theorem. Let us start
with the latter, which is not completely obvious since not all external
particles and anti-particles are renormalised with the same constant due to
the different absorptive parts.

The optical theorem asserts that
\begin{eqnarray}
\Gamma _{t}\sim \sum_{f}\int d\Pi _{f}\left| M\left( t^{\left( \hat{n}%
\right) }\left( p\right) \rightarrow f\right) \right| ^{2} &=&2\mathrm{Im}%
\left[ M\left( t^{\left( \hat{n}\right) }\left( p\right) \rightarrow
t^{\left( \hat{n}\right) }\left( p\right) \right) \right] \,,
\label{topdecay} \\
\Gamma _{\bar{t}}\sim \sum_{f}\int d\Pi _{f}\left| M\left( \bar{t}^{\left(
\hat{n}\right) }\left( p\right) \rightarrow f\right) \right| ^{2} &=&2%
\mathrm{Im}\left[ M\left( \bar{t}^{\left( \hat{n}\right) }\left( p\right)
\rightarrow \bar{t}^{\left( \hat{n}\right) }\left( p\right) \right) \right]
\,,  \label{antitopdecay}
\end{eqnarray}
where we have consider, just as an example, top ($t^{\left( \hat{n}\right)
}\left( p\right) $) and anti-top ($\bar{t}^{\left( \hat{n}\right) }\left(
p\right) $) decay, with $p$ and $\hat{n}$ being their momentum and
polarisation. Recalling that the incoming fermion and outgoing anti-fermion
spinors are renormalised with a common constant (see Eq. (\ref{fundamental}%
)) as are the outgoing fermion and incoming anti-fermion ones, it is
immediate to see that
\begin{eqnarray*}
M\left( t^{\left( \hat{n}\right) }\left( p\right) \rightarrow t^{\left( \hat{%
n}\right) }\left( p\right) \right) &=&\bar{u}^{\left( \hat{n}\right) }\left(
p\right) A_{33}\left( p\right) u^{\left( \hat{n}\right) }\left( p\right) \,,
\\
M\left( \bar{t}^{\left( \hat{n}\right) }\left( p\right) \rightarrow \bar{t}%
^{\left( \hat{n}\right) }\left( p\right) \right) &=&-\bar{v}^{\left( \hat{n}%
\right) }\left( p\right) A_{33}\left( -p\right) v^{\left( \hat{n}\right)
}\left( p\right) \,,
\end{eqnarray*}
where the minus sign comes from an interchange of two fermion operators and
where the subscripts in $A$ indicate family indices. Using the fact that
\begin{equation*}
u^{\left( \hat{n}\right) }\left( p\right) \otimes \bar{u}^{\left( \hat{n}%
\right) }\left( p\right) =\frac{\not{p}+m}{2m}\frac{1+\gamma ^{5}\not{n}}{2}%
\,,\qquad -v^{\left( \hat{n}\right) }\left( p\right) \otimes \bar{v}^{\left(
\hat{n}\right) }\left( p\right) =\frac{-\not{p}+m}{2m}\frac{1+\gamma ^{5}%
\not{n}}{2}\,,
\end{equation*}
with $n=\frac{1}{\sqrt{\left( p^{0}\right) ^{2}-\left( \vec{p}\cdot \hat{n}%
\right) ^{2}}}\left( \vec{p}\cdot \hat{n},p^{0}\hat{n}\right) $ being the
polarisation four-vector and performing some elementary manipulations we
obtain
\begin{eqnarray*}
\bar{u}^{\left( \hat{n}\right) }\left( p\right) A_{33}\left( p\right)
u^{\left( \hat{n}\right) }\left( p\right) &=&Tr\left[ \left( \frac{\not{p}+m%
}{2m}\frac{1+\gamma ^{5}\not{n}}{2}\right) \left( a\left( p^{2}\right) \not{p%
}L+b\left( p^{2}\right) \not{p}R+c\left( p^{2}\right) L+d\left( p^{2}\right)
R\right) \right] \\
&=&\frac{1}{4}Tr\left\{ \frac{\not{p}+m}{2m}\left[ \left( a\left(
p^{2}\right) +b\left( p^{2}\right) \right) \not{p}+c\left( p^{2}\right)
+d\left( p^{2}\right) \right] \right\} \\
&=&\frac{1}{4}Tr\left\{ \frac{-\not{p}+m}{2m}\left[ -\left( a\left(
p^{2}\right) +b\left( p^{2}\right) \right) \not{p}+c\left( p^{2}\right)
+d\left( p^{2}\right) \right] \right\} \\
&=&Tr\left[ \frac{-\not{p}+m}{2m}\frac{1+\gamma ^{5}\not{n}}{2}\left(
-a\left( p^{2}\right) \not{p}L-b\left( p^{2}\right) \not{p}R+c\left(
p^{2}\right) L+d\left( p^{2}\right) R\right) \right] \\
&=&-\bar{v}^{\left( \hat{n}\right) }\left( p\right) A_{33}\left( -p\right)
v^{\left( \hat{n}\right) }\left( p\right) \,,
\end{eqnarray*}
where we have decomposed $A_{33}\left( p\right) $ into its most general
Dirac structure. We thus conclude the equality between Eqs. (\ref{topdecay})
and (\ref{antitopdecay}) verifying that the lifetimes of top and anti-top are
identical. The detailed form of the wfr. constants, or whether they have
absorptive parts or not, does not play any role.

Even thought total decay widths for top and anti-top are identical the
partial ones need not to if $CP$ violation is present and some compensation
between different processes must take place. This issue is discussed in
detail in \cite{top}. Here we shall show that when $K=K^{\ast }$ the $CP$
invariance of the Lagrangian manifests itself in a zero asymmetry between
the partial differential decay rate of top and its $CP$ conjugate process.
The fact that the external renormalisation constants have dispersive parts
does not alter this conclusion. This is of course expected on rather general
grounds, so the following discussion has to be taken really as a
verification that no unexpected difficulties arise.

To illustrate this point let us consider the top decay channel $t\left(
p_{1}\right) \rightarrow W^{+}\left( p_{1}-p_{2}\right) +b\left(
p_{2}\right) $ and its $CP$ conjugate process $\bar{t}\left( \tilde{p}%
_{1}\right) \rightarrow W^{-}\left( \tilde{p}_{1}-\tilde{p}_{2}\right)
+b\left( \tilde{p}_{2}\right) .$ Let us note the respective amplitudes by $%
\mathcal{A}$ and $\mathcal{B}$ which are given as
\begin{eqnarray*}
\mathcal{A} &=&\varepsilon ^{\mu }\bar{u}^{\left( s_{2}\right) }\left(
p_{2}\right) A_{\mu }u^{\left( s_{1}\right) }\left( p_{1}\right) \,, \\
\mathcal{B} &=&-\tilde{\varepsilon}^{\mu }\bar{v}^{\left( s_{1}\right)
}\left( \tilde{p}_{1}\right) B_{\mu }v^{\left( s_{2}\right) }\left( \tilde{p}%
_{2}\right) \,,
\end{eqnarray*}
where $\tilde{a}^{\mu }=a_{\mu }=\left( a^{0},-a^{i}\right) $ for any
four-vector. Considering contributions up to including next-to-leading
corrections we have
\begin{eqnarray*}
A_{\mu } &=&-i\frac{e}{\sqrt{2}s_{W}}\left[ \left( \bar{Z}^{\frac{1}{2}%
bL}K^{\dagger }Z^{\frac{1}{2}tL}+K^{\dagger }\delta _{V}+\delta K^{\dagger
}\right) \gamma _{\mu }L+\delta F_{\mu }\right] \,, \\
B_{\mu } &=&-i\frac{e}{\sqrt{2}s_{W}}\left[ \left( \bar{Z}^{\frac{1}{2}%
tL}KZ^{\frac{1}{2}bL}+K\delta _{V}+\delta K\right) \gamma _{\mu }L+\delta
G_{\mu }\right] \,,
\end{eqnarray*}
with $\delta _{V}$ $=\frac{\delta e}{e}-\frac{\delta s_{W}}{s_{W}}+\frac{1}{2%
}\delta Z_{W}$ and $\delta F_{\mu }$ and $\delta G_{\mu }$ are given by the
one-loop diagrams. From a direct computation it can be seen that if $%
K=K^{\ast }$ this implies

\begin{equation}
\bar{Z}^{\frac{1}{2}L}=\left( Z^{\frac{1}{2}L}\right) ^{T}\,,\quad \bar{Z}^{%
\frac{1}{2}R}=\left( Z^{\frac{1}{2}R}\right) ^{T}\,,\quad \tilde{\varepsilon}%
^{\mu }\delta G_{\mu }=\varepsilon ^{\mu }\gamma ^{2}\delta F_{\mu
}^{T}\gamma ^{2}\,,  \label{CPfacts}
\end{equation}
where the superscript $T$ means transposition with respect to all indices
(family indices in the case of $Z^{\frac{1}{2}L}$ and $Z^{\frac{1}{2}R}$ and
Dirac indices in the case of $\delta F_{\mu }$ ). Using
\begin{equation*}
i\gamma ^{2}\bar{u}^{\left( s\right) T}\left( p\right) =sv^{\left( s\right)
}\left( \tilde{p}\right) \,,\qquad u^{\left( s\right) T}\left( p\right)
i\gamma ^{2}=-s\bar{v}^{\left( s\right) }\left( \tilde{p}\right) \,,
\end{equation*}
where $s=\pm 1$, depending on the spin direction in the $\hat{z}$ axis, we
obtain
\begin{eqnarray*}
\mathcal{A} &=&\frac{-ie}{\sqrt{2}s_{W}}\varepsilon ^{\mu }\bar{u}^{\left(
s_{2}\right) }\left( p_{2}\right) \left[ \left( \bar{Z}^{\frac{1}{2}%
bL}K^{\dagger }Z^{\frac{1}{2}tL}+K^{\dagger }\delta _{V}+\delta K^{\dagger
}\right) \gamma _{\mu }L+\delta F_{\mu }\right] u^{\left( s_{1}\right)
}\left( p_{1}\right) \\
&=&\frac{-ie}{\sqrt{2}s_{W}}\varepsilon ^{\mu }u^{\left( s_{1}\right)
T}\left( p_{1}\right) \left[ L\left( \left( Z^{\frac{1}{2}tL}\right)
^{T}K^{\ast }\left( \bar{Z}^{\frac{1}{2}bL}\right) ^{T}+K^{\ast }\delta
_{V}+\delta K^{\ast }\right) \gamma _{\mu }^{T}+\delta F_{\mu }^{T}\right]
\bar{u}^{\left( s_{2}\right) T}\left( p_{2}\right) \\
&=&\frac{-s_{1}s_{2}ie}{\sqrt{2}s_{W}}\varepsilon ^{\mu }\bar{v}^{\left(
s_{1}\right) }\left( \tilde{p}_{1}\right) \gamma ^{2}\left[ L\left( \left(
Z^{\frac{1}{2}tL}\right) ^{T}K^{\ast }\left( \bar{Z}^{\frac{1}{2}bL}\right)
^{T}+K^{\ast }\delta _{V}+\delta K^{\ast }\right) \gamma _{\mu }^{T}+\delta
F_{\mu }^{T}\right] \gamma ^{2}v^{\left( s_{2}\right) }\left( \tilde{p}%
_{2}\right) \\
&=&\frac{-s_{1}s_{2}ie}{\sqrt{2}s_{W}}\varepsilon ^{\mu }\bar{v}^{\left(
s_{1}\right) }\left( \tilde{p}_{1}\right) \left[ \left( \left( Z^{\frac{1}{2}%
tL}\right) ^{T}K^{\ast }\left( \bar{Z}^{\frac{1}{2}bL}\right) ^{T}+K^{\ast
}\delta _{V}+\delta K^{\ast }\right) \gamma _{\mu }^{\dagger }L+\gamma
^{2}\delta F_{\mu }^{T}\gamma ^{2}\right] v^{\left( s_{2}\right) }\left(
\tilde{p}_{2}\right) \,,
\end{eqnarray*}
now using Eq. (\ref{CPfacts}) we see that if no $CP$ violating phases are
present in the CKM matrix $K$ (and therefore neither in $\delta K,$ Eq. (\ref
{deltaK})) we obtain that $\mathcal{A}=-s_{1}s_{2}\mathcal{B}$ and thus
\begin{equation*}
\left| \mathcal{A}\right| ^{2}=\left| \mathcal{B}\right| ^{2}\,.
\end{equation*}

Note again that when $CP$ violating phases are present we can expect in
general non-vanishing phase-space dependent asymmetries for the different
channels. Once we sum over all channels and integrate over the final state
phase space a compensation must take place as we have seen guaranteed by
unitarity and $CPT$ invariance. Using a set of wfr. constants with
absorptive parts as advocated here (and required by gauge invariance) leads
to different results than using the prescription originally advocated in
\cite{Denner}, in particular using Eq. (\ref{res2}) for $K\neq K^{\ast }$ we
expect $\Delta _{3j}^{\left( t~decay\right) }-\Delta _{3j}^{\left( \bar{t}%
~decay\right) }\neq 0$.

\section{Conclusions}

\label{conc}

Let us recapitulate our main results. We hope, first of all, to have
convinced the reader that \emph{there is} a problem with what appears to be
the commonly accepted prescription for dealing with wave function
renormalisation when mixing is present. The situation is even further
complicated by the appearance of $CP$ violating phases. The problem has a
twofold aspect. On the one hand the prescription of \cite{Denner} does not
diagonalise the propagator matrix in flavour space in what respects to the
absorptive parts. On the other hand it yields gauge \emph{dependent}
amplitudes, albeit gauge \emph{independent} modulus squared of the
amplitudes. This is not satisfactory: interference with e.g. strong phases
may reveal an unacceptable gauge dependence.

The only solution is to accept wfr. constants that do not satisfy a
pseudo-hermiticity condition due to the presence of the absorptive parts,
which are neglected in \cite{Denner}. This immediately brings about some
gauge \emph{independent} absorptive parts which appear even in the modulus
squared amplitude and which are neglected in the treatment of \cite{Denner}.
Furthermore, these parts (and the gauge dependent ones) cannot be absorbed
in unitary redefinitions of the CKM matrix which are the only ones allowed
by Ward identities. We have checked that ---although unconventional--- the
presence of the absorptive parts in the wfr. constants is perfectly
compatible with basic tenets of field theory and the Standard Model.
Numerically we have found the differences to be important, at the order of
the half per cent. Small, but relevant in the future. This information will
be relevant to extract the experimental values of the CKM mixing matrix.

Traditionally, wave function renormalisation seems to have been the ``poor
relative'' in the Standard Model renormalisation program. We have seen here
that it is important on two counts. First because it is related to the
counter terms for the CKM mixing matrix, although the on-shell values for
wave function constants cannot be directly used there. Second because they
are crucial to obtain gauge independent $S$ matrix elements and observables.
While using our wfr. constants (but not the ones in \cite{Denner}) for the
external legs is strictly equivalent to considering reducible diagrams (with
on-shell mass counter terms) the former procedure is considerably more
practical.

\section*{Acknowledgements}

This work was triggered by many fruitful discussions with B. A. Kniehl
concerning previous work of two of the present authors. We thank him for
suggestions, criticism and for a careful reading of a preliminary draft. The
work of D.E. and P.T. is supported in part by TMR, EC--Contract No.
ERBFMRX--CT980169 (EURODAPHNE) and contracts MCyT FPA2001-3598 and CIRIT
2001SGR-00065 as well as by a Distinci\'{o} from the Generalitat de
Catalunya. J.M. acknowledges a fellowship from Generalitat de Catalunya,
grant 1998FI-00614.


\begin{thebibliography}{99}
\bibitem{Amato:1998xt}  S.~Amato \textit{et al.} [LHCb Collaboration],
CERN-LHCC-98-4.


\bibitem{Pich:1997ga}  A.~Pich, 
arXiv:hep-ph/9711279. 

\bibitem{Balzereit:1999id}  C.~Balzereit, T.~Mannel and B.~Plumper,
Eur.\ Phys.\ J.\ C \textbf{9}, 197 (1999) [arXiv:hep-ph/9810350].

\bibitem{Hollik}  W.F.L. Hollik, Fortsch.\ Phys.\ 38 (1990) 165-260,
Fortsch.\ Phys.\ \textbf{34} (1986) 687-751.

\bibitem{Aoki}  K.I. Aoki, Z. Hioki, M. Konuma, R. Kawabe, T. Muta, Prog.\
Theor.\ Phys.\ Suppl. \textbf{73} (1982) 1-225.

\bibitem{DennerSack}  A. Denner and T. Sack,
Nucl.\ Phys.\ B \textbf{347} (1990) 203. 

\bibitem{Denner}  A. Denner, Fortsch. Phys. \textbf{41} (1993) 4, 307-420.

\bibitem{Grassi}  P. Gambino, P.A. Grassi, F. Madricardo, Phys.\ Lett.\ B
\textbf{454} (1999) 98-104.

\bibitem{Diener:2001qt}  K.~P.~Diener and B.~A.~Kniehl,
Nucl.\ Phys.\ B \textbf{617}, 291 (2001) [arXiv:hep-ph/0109110].

\bibitem{Barroso}  A. Barroso, L. Brucher and R. Santos,
Phys.\ Rev.\ D \textbf{62} (2000) 096003 [arXiv:hep-ph/0004136].


\bibitem{Yamada}  Y. Yamada,
Phys.\ Rev.\ D \textbf{64} (2001) 036008 [arXiv:hep-ph/0103046].

\bibitem{Kniehl}  B. A. Kniehl, F. Madricardo, M. Steinhauser, Phys. Rev. D
\textbf{62} (2000) 073010.

\bibitem{espriumanz}  D.~Espriu and J.~Manzano,
Phys.\ Rev.\ D \textbf{63}, 073008 (2001) [arXiv:hep-ph/0011036].

\bibitem{Sirlin:1991fd}  A. Sirlin,
Phys.\ Rev.\ Lett.\ \textbf{67} (1991) 2127. 

\bibitem{Sirlin:1998dz}  A. Sirlin,
arXiv:hep-ph/9811467. 

\bibitem{Grassi:2001dz}  P. A. Grassi, B. A. Kniehl and A. Sirlin,
Phys.\ Rev.\ Lett.\ \textbf{86} (2001) 389 [arXiv:hep-th/0005149].

\bibitem{Grassi:2001bz}  P. A. Grassi, B. A. Kniehl and A. Sirlin,
arXiv:hep-ph/0109228. 

\bibitem{Marciano}  W. J. Marciano and A. Sirlin,
Phys.\ Rev.\ D \textbf{22} (1980) 2695 [Erratum-ibid.\ D \textbf{31} (1980)
213]. 

\bibitem{Gambino}  P. Gambino, P. A. Grassi, Phys.\ Rev.\ D \textbf{62}
(2000) 076002.

\bibitem{Nielsen}  N. K. Nielsen, Nucl. Phys. B \textbf{101}, (1975) 173.

\bibitem{Piguet}  O. Piguet, K. Sibold, Nucl. Phys. B \textbf{253} (1985)
517-540.

\bibitem{Sibold}  E. Kraus, K. Sibold, hep-th/9406115.

\bibitem{top}  D.~Atwood, S.~Bar-Shalom, G.~Eilam and A.~Soni,
Phys.\ Rept.\ \textbf{347}, 1 (2001) [arXiv:hep-ph/0006032].
\end{thebibliography}
\end{document}